\documentclass[10pt,a4paper]{article}
\usepackage{jheppub,tcolorbox}
\usepackage[T1]{fontenc}
\usepackage[utf8]{inputenc}
\usepackage{amsmath,amssymb}

\DeclareMathOperator{\Tr}{Tr}

\DeclareUnicodeCharacter{221E}{\infty}

\title{\boldmath{The Off-Shell Recursion for Gravity \\ and the Classical Double Copy for currents}}

\author[a]{Kyoungho Cho}
\author[b]{Kwangeon Kim}
\author[a,c]{Kanghoon Lee}

\affiliation[a]{Asia Pacific Center for Theoretical Physics, Postech, Pohang 37673, Korea}
\affiliation[b]{Department of Physics, Yonsei University, Seoul 03722, Korea}
\affiliation[c]{Department of Physics, Postech, Pohang 37673, Korea}

\emailAdd{kyoungho.cho@apctp.org}
\emailAdd{kim64656@yonsei.ac.kr}
\emailAdd{kanghoon.lee1@gmail.com}

\abstract{
We construct the off-shell recursion for gravity and the graviton current for the perturbative double field theory (DFT). We first formulate the perturbative DFT, which is equivalent but simpler to perturbative general relativity, to all-orders in fluctuations of generalised metric. The perturbative action and equations of motion (EoM) are derived to arbitrary order for pure gravity case. We then derive the graviton off-shell recursion, the gravity counterpart of the Berends-Giele recursion in Yang-Mills theory, through the so-called perturbiner method using the EoM of the perturbative DFT. We solve the recursion iteratively and obtain the graviton off-shell currents explicitly. We then discuss the classical double copy for the off-shell currents. We present the current KLT relation for gravity by extending the result proposed by Mizera and Skrzypek for the non-gravitational effective field theories. The relation represents graviton currents by squaring gluon currents with the KLT kernel up to gauge transformation and regular terms that do not have any pole. Finally we discuss the off-shell conservation of currents for nonlinear gauge choices.
}

\begin{document}

\preprint{APCTP Pre2021 - 019}
\maketitle

\section{\label{sec:1}Introduction}

The perturbation theory of general relativity (GR) is one of the most successful techniques of gravity. The recent observations, such as the gravitational wave detection \cite{LIGOScientific:2016aoc,LIGOScientific:2017vwq} and cosmic microwave background \cite{Planck:2018vyg,Planck:2018jri}, confirmed the validity of the perturbative GR. However, the perturbative GR usually involves too many calculations both in the classical and the quantum levels. The main source of such complications is infinite expansions of the metric perturbation from the square root of the metric determinant $\sqrt{-g}$ and the inverse metric $g^{-1}$. These generate infinitely many irregular Feynman vertices that we cannot fully handle. In particular the calculation of graviton scattering amplitude through the traditional approach using the Feynman diagram is limited despite its importance. Therefore, the development of tools that simplifies the computation is important in perturbative gravity.

As an alternative theory of gravity, double field theory (DFT) reformulates the closed string low energy effective field theory by requiring the manifest $O(D,D)$ T-duality \cite{Siegel:1993bj,Siegel:1993th,Siegel:1993xq,Hull:2009mi,Hull:2009zb,Hohm:2010jy,Hohm:2010pp}. It provides a geometric framework for the entire massless NS-NS sector encoded in DFT field variables: the generalised metric and the DFT dilaton. DFT is equivalent to GR but more strongly constrained by the $O(D,D)$ structure, and the doubled local Lorentz groups, which cannot be seen in GR directly. In contrast to GR, the inverse and the square root of the determinant of the generalised metric are trivial, and these do not appear in the DFT action. These structures simplify the calculations in DFT compare to GR. For instance, the supersymmetric DFT has a more straightforward SUSY structure than the conventional supergravities \cite{Jeon:2011sq,Jeon:2012hp,Jeon:2011vx,Hohm:2011nu,Butter:2021dtu}. 

In the first part of this work, we construct the general framework of the perturbative DFT to all orders in fluctuation. This framework extends the previous studies of the finite order perturbation theory of DFT up to cubic orders \cite{Ko:2015rha,Hohm:2015ugy}. We solve the perturbative $\mathit{O}(D,D)$ constraint exactly and analyse the structure of the perturbed generalised metric according to the chirality for the background projection operators. The graviton field is identified in the mixed chirality sector, and the chiral and the antichiral sectors are composite fields of the graviton fields. We further find the exact map between the perturbative generalised metric in DFT and the perturbative metric in GR. This map shows that the perturbative DFT is the consistent theory of perturbative gravity. We introduce a gauge fixing condition, which yields the same with the de Donder gauge condition and derive the perturbative EoM to arbitrary order in perturbation.

In addition to being a powerful calculation tool, a remarkable aspect of DFT is the relation with the double copy. The double copy is a relation between gravity and the Yang-Mills theory, which originated in graviton scattering, and it defines a map between graviton and gluon amplitudes \cite{Kawai:1985xq,Bern:2008qj,Bern:2010ue,BjerrumBohr:2009rd,Stieberger:2009hq,Bern:2010yg,BjerrumBohr:2010zs,Feng:2010my,Tye:2010dd,Mafra:2011kj,Monteiro:2011pc,BjerrumBohr:2012mg}. It turns out that DFT and the double copy share the common origin from the closed string theory, in particular the factorisation into left- and right-moving sectors \cite{Hohm:2011dz,Cheung:2016say,Cheung:2017kzx,Lee:2018gxc,Cho:2019ype,Kim:2019jwm,Angus:2021zhy,Lescano:2020nve,Diaz-Jaramillo:2021wtl}. The double copy prescription has recently been extended to solutions of the classical equations of motion (EoM), the so-called classical double copy \cite{Monteiro:2014cda,Luna:2015paa,Luna:2016due,Lee:2018gxc}, by establishing the map between the gravity and gauge theory solutions \cite{Anastasiou:2014qba,Borsten:2015pla,Anastasiou:2016csv,Goldberger:2016iau,Luna:2016hge,Goldberger:2017frp,Cardoso:2016ngt,Anastasiou:2018rdx,Shen:2018ebu,CarrilloGonzalez:2018ejf,Plefka:2018dpa,Berman:2018hwd,Gurses:2018ckx,Luna:2018dpt,Alawadhi:2019urr,Kim:2019jwm,Banerjee:2019saj,Bahjat-Abbas:2020cyb,Alfonsi:2020lub,Keeler:2020rcv,Elor:2020nqe,Berman:2020xvs,Momeni:2020vvr,Alawadhi:2020jrv,Godazgar:2020zbv,Prabhu:2020avf,Carrasco:2020ywq,Ferrero:2020vww,Chacon:2020fmr,White:2020sfn,Monteiro:2020plf,Lescano:2021ooe,Alkac:2021bav,Lescano:2021but,Campiglia:2021srh,Chacon:2021wbr,Farnsworth:2021wvs,Alkac:2021seh}. However, despite the remarkable developments, there is no known algorithmic prescription for understanding the relation between the single/double copy maps of classical solutions and the KLT relation for the tree-level scattering amplitudes. 

To address this issue we exploit the so-called \textit{perturbiner method} by Rosly and Selivanov \cite{Rosly:1996vr,Rosly:1997ap,Selivanov:1997aq,Selivanov:1997an,Selivanov:1997ts} that provides a direct connection between solutions of EoM and tree-level scattering amplitudes. The perturbiner expansion is a generating function of the Berends-Giele currents in Yang-Mills theory, which gives an efficient tool for computing the gluon scattering amplitudes \footnote{See also \cite{Macrelli:2019afx,Arvanitakis:2019ald,Lopez-Arcos:2019hvg,Jurco:2019yfd,Arvanitakis:2020rrk,Gomez:2020vat} for the $L_\infty$ algebra in the BG recursion relation.}. Berends-Giele constructed the recursion relation \cite{Berends:1987me} in terms of the off-shell currents using the recursive structure of the Feynman vertices of YM theory. On the other hand, the perturbiner method generates the same recursion relation by applying the perturbiner expansion to the classical equations of motion (EoM). This method has been applied to some non-gravitational effective field theories successfully \cite{Mafra:2016ltu,Mizera:2018jbh,Garozzo:2018uzj}. Thus it is natural to seek the off-shell recursion relation for gravity. However, the structure of the perturbative GR is completely different from the Yang-Mills theory. As we have stated, infinitely many irregular Feynman vertices arise in the perturbative GR, and it makes difficult to find the recursive structure. Therefore, the perturbiner method that does not use the Feynman vertices is suitable in developing the off-shell recursion for gravity. Recently, there have been several works deriving the graviton off-shell recursions from the perturbative GR \cite{Cheung:2016say,Cheung:2017kzx,Gomez:2021shh,Cheung:2021zvb}, and the associated graviton off-shell currents have been obtained using the recursions.

The second part of this work constructs the graviton off-shell recursion through the perturbiner method for the perturbative DFT. Exploiting the spinor-helicity formalism, we obtain the explicit graviton off-shell currents by solving the recursion iteratively. Remarkably, the graviton currents exhibit the current KLT relation, representing the graviton currents by the square of gluon currents. Recently, the current KLT relation for the non-gravitational effective theories are studied in \cite{Mizera:2018jbh}, and the color-kinematic duality for the off-shell currents is introduced in \cite{Cheung:2021zvb}. Comparing with the non-gravitational current KLT relation, the gravitational case contains contributions from the regular terms for the propagators of the off-shell leg, which do not contribute to the scattering amplitudes. Since the perturbiner expansions are the solutions of the EoM, the current KLT relation gives another viewpoint on the perturbative classical double copy. As a consistency check, we examine the conservation of our graviton off-shell currents. First, we demonstrate that the currents are conserved up to gauge fixing conditions when the gauge choice is nonlinear in fields. To support this, we check the off-shell conservation of gluon currents under the Gervais-Neveu gauge condition, which is quadratic in the gauge field. The conservation equation for the graviton currents yields the gauge condition of the generalised metric, and it confirms our results.

The structure of this paper is as follows. In section \ref{Sec:2} we will construct the perturbation theory of DFT. We will analyze the properties of the perturbed generalised metric and compare with the usual metric perturbation. We will consider the equations of motion for the perturbed generalised metric which corresponds to the perturbed Einstein equation. In section \ref{Sec:3}, we will review the perturbiner method for the Yang-Mills theory and then apply to the DFT. We will introduce the perturbiner expansion for the graviton field and construct the off-shell recursion relation for graviton off-shell currents by substituting the expansion into the perturbative DFT EoM. In section \ref{Sec:4} we will solve the graviton recursion relation iteratively and  show the current KLT relation. Section \ref{Sec:5} discuss the conservation of currents in nonlinear gauge fixing. Section \ref{Sec:6} gives our conclusion.

\section{\label{Sec:2}Perturbation of Double Field Theory}

In this section, we construct the perturbation theory of double field theory (DFT) as an alternative tool to the perturbative GR. We first analyse the structure of the perturbation of the generalised metric by solving the $\mathit{O}(D,D)$ constraint exactly. We find the relation between the generalised metric perturbation and the metric perturbation in GR. Next, we construct action and EoM of the perturbative DFT for pure gravity including all-orders in the generalised metric perturbation. We also discuss the gauge choice of the generalised metric, which is closely related to the de Donder gauge fixing in perturbative GR.

\subsection{Perturbation of the generalised metric}
Double field theory is a reformulation of low energy effective field theory of the closed string theory that is manifest under $\mathit{O}(D,D)$ T-duality. The field content of DFT is given by $\mathit{O}(D,D)$ tensors due to the manifest $\mathit{O}(D,D)$ covariance: the generalised metric $\mathcal{H}_{MN}$ and DFT dilaton $d$. The generalised metric is not only an arbitrary symmetric $2D\times 2D$ matrix but also constrained by the $\mathit{O}(D,D)$ constraint, 
\begin{equation}
  \mathcal{H}_{MN} \mathcal{J}^{NP} \mathcal{H}_{PQ} = \mathcal{J}_{MQ}\,,
\label{ODD_const}\end{equation}
where $\mathcal{J}_{MN}$ is the $\mathit{O}(D,D)$ metric parametrised as
\begin{equation}
  \mathcal{J}_{MN} = \begin{pmatrix} 0 & \delta^{\mu}{}_{\nu} \\ \delta_{\mu}{}^{\nu} & 0\end{pmatrix}\,.
\label{}\end{equation}
Note that $\mathcal{J}_{MN}$ defines the inner product for the doubled tangent space instead of $\mathcal{H}_{MN}$. Thus we should raise and lower the $\mathit{O}(D,D)$ vector indices using $\mathcal{J}^{MN}$ and $\mathcal{J}_{MN}$ respectively. 

If we contract $\mathcal{J}^{MN}$ with \eqref{ODD_const}, the $\mathit{O}(D,D)$ constraint can be recast by
\begin{equation}
  \mathcal{H}_{M}{}^{N} \mathcal{H}_{N}{}^{P} = \delta_{M}{}^{P}\,,
\label{}\end{equation}
which means that the inverse of the generalised metric $\mathcal{H}_{M}{}^{N}$ is itself. This property ensures to define a pair of projection operators $P_{M}{}^{N}$ and $\bar{P}_{M}{}^{N}$ in terms of $\mathcal{H}$
\begin{equation}
  P_{M}{}^{N} = \frac{1}{2} \Big(\delta_{M}{}^{N} + \mathcal{H}_{M}{}^{N}\Big)\,, 
  \qquad 
  \bar{P}_{M}{}^{N} = \frac{1}{2} \Big(\delta_{M}{}^{N} - \mathcal{H}_{M}{}^{N}\Big)\,.
\label{def_PPb}\end{equation}
These satisfy the standard properties of projection operators, $P^{2} = P$ and $\bar{P}^{2}=\bar{P}$ and $P \bar{P} = \bar{P} P = 0$. We can also define chiralities associated with the projection operators. The chiral and antichiral sectors of the doubled tangent space are associated with the $P$ and $\bar{P}$ respectively. 

The characteristic feature of DFT related with the double copy is the double local Lorentz group
\begin{equation}
  \mathit{O}(1,D-1)_{L}\times\mathit{O}(1,D-1)_{R} \,,
\label{double_local_Lorentz}\end{equation}
which is a Lorentzian version of the maximal compact subgroup of $\mathit{O}(D,D)$. In the closed string point of view, it arises from the left-right mover decomposition of the closed-string mode expansion and shares a common origin with the KLT relation \cite{Kawai:1985xq}. This structure leads to the double-vielbein or generalised frame fields, $V_{M}{}^{m}$ and $\bar{V}_{M}{}^{\bar{m}}$, where $m$ and $\bar{m}$ are local frame indices associated with the double Lorentz group, $\mathit{O}(1,D-1)_{L}$ and $\mathit{O}(1,D-1)_{R}$ respectively \cite{Jeon:2010rw,Jeon:2011cn}. These satisfy the defining conditions
\begin{equation}
  V_{Mm} \eta^{mn}(V^{t})_{nN}= P_{MN}\,, \qquad \bar{V}_{M\bar{m}} \bar{\eta}^{\bar{m}\bar{n}}(\bar{V}^{t})_{\hat{\bar{n}}\hat{N}} =-\bar{P}_{MN}\,,
\label{defining_VVbar}\end{equation}
where $\eta^{mn} = \bar{\eta}^{\bar{m}\bar{n}} = \text{diag}(-1,1,\cdots,1) $.

We can parametrise $\mathcal{H}$ in terms of the massless NS-NS fields, metric $g_{\mu\nu}$ and the antisymmetric two-form $B_{\mu\nu}$ 
\begin{equation}
  \mathcal{H} = \begin{pmatrix} g^{\mu\nu} & - g^{\mu\rho} B_{\rho\nu} \\ B_{\mu\rho} g^{\rho\nu} & g_{\mu\nu} - B_{\mu\rho} g^{\rho\sigma} B_{\sigma\nu} \end{pmatrix}\,. 
\label{para_H}\end{equation}
Further, the DFT dilaton $d$, which plays a role of  is represented by
\begin{equation}
  e^{-2d} = \sqrt{-g}e^{-2\phi}\,.
\label{DFT_dilaton}\end{equation}
Similarly, the parametrization of the double-vielbeins are given by
\begin{equation}
  V_{M}{}^{m} = \frac{1}{\sqrt{2}}\begin{pmatrix} (e^{-1})^{\mu m} \\ e_{\mu}{}^{m} \end{pmatrix} \,, 
  \qquad 
  \bar{V}_{M}{}^{\bar{m}} = \frac{1}{\sqrt{2}}\begin{pmatrix} (\bar{e}^{-1})^{\mu \bar{m}} \\ -\bar{e}_{\mu}{}^{\bar{m}} \end{pmatrix}\,,
\label{double_vielbein}\end{equation}
where $e_{\mu}{}^{m}$ and $\bar{e}_{\mu}{}^{\bar{m}}$ are local frame fields associated with the same metric, $g_{\mu\nu} = e_{\mu}{}^{m} e_{\mu}{}^{n}\eta_{mn} = \bar{e}_{\mu}{}^{\bar{m}} \bar{e}_{\mu}{}^{\bar{n}}\bar{\eta}_{\bar{m}\bar{n}}$. These are related by the local Lorentz group.

Let us consider the perturbative expansion of the generalised metric around a flat background $\mathcal{H}_{0MN}$
\begin{equation}
  \mathcal{H}_{MN} = \mathcal{H}_{0MN} + \kappa\Pi_{MN}\,, \qquad \mathcal{H}_{0MN} = \begin{pmatrix} \eta^{\mu\nu} & 0 \\ 0 &\eta_{\mu\nu} \end{pmatrix}\,,
\label{H_pert1}\end{equation}
where $\Pi$ is the fluctuation of the generalised metric and $\kappa$ is an expansion parameter and $\eta_{\mu\nu}=\text{diag}(-1,1,1,\cdots,1) $. Introducing background projection operators with respect to $\mathcal{H}_{0}$ using \eqref{def_PPb}, $P_{0} = \frac{1}{2} \big(\delta + \mathcal{H}_{0}\big)$ and $\bar{P}_{0} = \frac{1}{2} \big(\delta - \mathcal{H}_{0}\big)$, we decompose the fluctuation $\Pi$ into four pieces according to the background chiralities
\begin{equation}
\begin{aligned}
  P_{0M}{}^{P} P_{0N}{}^{Q} \Pi_{PQ}  = \Delta_{MN}\,, &\quad P_{0M}{}^{P} \bar{P}_{0N}{}^{Q} \Pi_{PQ} = \mathcal{E}_{MN} \,, 
  \\ 
  \bar{P}_{0M}{}^{P} P_{0N}{}^{Q} \Pi_{PQ} = \bar{\mathcal{E}}_{MN} \,, &\quad \bar{P}_{0M}{}^{P} \bar{P}_{0N}{}^{Q}\Pi_{PQ} = \bar{\Delta}_{MN}\,.
\end{aligned}\label{decompos_PI}
\end{equation}
Here $\Delta$ and $\bar{\Delta}$ are symmetric, and $\bar{\mathcal{E}} = \mathcal{E}^{t}$. Then the perturbation of generalised metric \eqref{H_pert1} is rewritten by
\begin{equation}
  \mathcal{H}_{MN} = \mathcal{H}_{0MN} + \kappa\Big(\mathcal{E}_{MN} + \bar{\mathcal{E}}_{MN} + \Delta_{MN} + \bar{\Delta}_{MN}\Big)\,.
\label{perturbed_H}\end{equation}

The fluctuations $\mathcal{E}$, $\bar{\mathcal{E}}$, $\Delta$ and $\bar{\Delta}$ are not arbitrary, but constrained by the $\mathit{O}(D,D)$ constraint \eqref{ODD_const}. One may decompose the constraint according to the background chiralities as \eqref{decompos_PI}
\begin{equation}
\begin{aligned}
  2 \Delta_{MN} + \Delta_{M}{}^{P} \Delta_{PN} + \kappa^{2} \mathcal{E}_{M}{}^{P} \bar{\mathcal{E}}_{PN}= 0\,,
  \\
  -2 \bar{\Delta}_{MN}  + \bar{\Delta}_{M}{}^{P} \bar{\Delta}_{PN} +\kappa^{2}\bar{\mathcal{E}}_{M}{}^{P} \mathcal{E}_{PN} = 0\,,
  \\
  \kappa\Delta_{M}{}^{P} \mathcal{E}_{PN} + \kappa\mathcal{E}_{M}{}^{P} \bar{\Delta}_{PN}= 0 \,,
  \\ 
  \kappa \bar{\Delta}_{M}{}^{P} \bar{\mathcal{E}}_{PN} + \kappa \bar{\mathcal{E}}_{M}{}^{P} \Delta_{PN}=0\,,
\end{aligned}\label{odd_component}
\end{equation}
where we have absorbed $\kappa$ into $\Delta$ and $\bar{\Delta}$ for convenience. One can solve these constraints perturbatively in $\kappa$. The first and the second equations imply that $\Delta $ and $\bar{\Delta}$ cannot be a fundamental degrees of freedom, but functions of $\mathcal{E}$ and $\bar{\mathcal{E}}$. In other words, $\mathcal{E}$ is the linear perturbation of $\mathcal{H}$.\footnote{Of course $\mathcal{E}$ does not have to be linear perturbation only. It may be expanded further by
\begin{equation}
  \kappa\mathcal{E} = \kappa\mathcal{E}^{{\scriptscriptstyle(2)}} + \kappa^{2} \mathcal{E}^{{\scriptscriptstyle(2)}} + \kappa^{3} \mathcal{E}^{\scriptscriptstyle(3)} + \cdots \,.
\label{}\end{equation}
The higher order terms $\mathcal{E}^{{\scriptscriptstyle(n)}}$ where $n>1$ corresponds to the higher order perturbations of metric in GR. In this paper, we will focus on the linear perturbation and identify $\mathcal{E} = \mathcal{E}^{{\scriptscriptstyle(1)}}$.}
The third and fourth equations imply that $\Delta$ and $\bar{\Delta}$ are expanded by the product between $\mathcal{E}$ and $\bar{\mathcal{E}}$
\begin{equation}
\begin{aligned}
    \Delta_{MN} = \sum_{n=1} a_{n} \big(\kappa^{2}\mathcal{E}\bar{\mathcal{E}}\big)^{n}{}_{MN} \,, \qquad \bar{\Delta}_{MN} = \sum_{n=1} b_{n} \big(\kappa^{2}\bar{\mathcal{E}}\mathcal{E}\big)^{n}{}_{MN}\,.
\end{aligned}\label{}
\end{equation}
and the coefficients of the series are related by $a_{n} = -b_{n}$. If we substitute these results into the first and second equations, we have the following recursion relation and initial condition
\begin{equation}
\begin{aligned}
  &a_{n+1} +\frac{1}{2} \sum_{p=1}^{n} a_{n+1-p}\cdot a_{p} = 0\,, \quad
  \mbox{for} ~n>1
  \\
  &a_{1} = -\frac{1}{2}\,.
\end{aligned}\label{}
\end{equation}
There is an exact solution for the recurrence relation given by the binomial expansion
\begin{equation}
  a_{n} = (-1)^{n}{\frac{1}{2}\choose n} \quad \mbox{for} ~n\geq1\,,
\label{}\end{equation}
which is nothing but the coefficients of the Taylor expansion of the square root. Then $\Delta$ and $\bar{\Delta}$ are represented by 
\begin{equation}
\begin{aligned}
  \Delta_{MN} &= \Big(-\mathbf{1}+\sqrt{\mathbf{1} -\kappa^{2}\mathcal{E}\bar{\mathcal{E}}}\Big){}_{MN}\,, 
  \\ 
  \bar{\Delta}_{MN} &= \Big(\mathbf{1}-\sqrt{\mathbf{1} -\kappa^{2}\bar{\mathcal{E}}\mathcal{E}}\Big){}_{MN}\,,
\end{aligned}\label{D_Db_expansion}
\end{equation}
and the first few terms are
\begin{equation}
\begin{aligned}
    \Delta_{MN} &= - \frac{1}{2} \big(\kappa^{2}\mathcal{E}\bar{\mathcal{E}}\big)_{MN} - \frac{1}{8} \big(\kappa^{2}\mathcal{E}\bar{\mathcal{E}}\big)^{2}_{MN} - \frac{1}{16} \big(\kappa^{2}\mathcal{E}\bar{\mathcal{E}}\big)^{3}_{MN} -\cdots
    \\
    \bar{\Delta}_{MN} &= \frac{1}{2} \big(\kappa^{2}\bar{\mathcal{E}}\mathcal{E}\big)_{MN} + \frac{1}{8} \big(\kappa^{2}\bar{\mathcal{E}}\mathcal{E}\big)^{2}_{MN} + \frac{1}{16} \big(\kappa^{2}\bar{\mathcal{E}}\mathcal{E}\big)^{3}_{MN} +\cdots\,.
\end{aligned}\label{}
\end{equation}

Note that the components of the $\mathit{O}(D,D)$ vector indices are not independent and have redundancies. To remove such redundancies, we introduce a pair of projection operators that project out the $\mathit{O}(D,D)$ vector indices to $D$-dimensional undoubled vector indices by using projectors. Let us introduce a pair of projectors $\Theta_{\mu}{}^{M}$ and $\bar{\Theta}_{\mu}{}^{M}$ defined in terms of the background double-vielbeins, $V_{0}$ and $\bar{V}_{0}$ \eqref{double_vielbein} 
\begin{equation}
  \Theta_{\mu}{}^{M} = e_{0\mu}{}^{m} \big(V_{0}^{t}\big)_{m}{}^{M} = \frac{1}{\sqrt{2}} \begin{pmatrix} \delta_{\mu}{}^{\nu}\\ \eta_{\mu\nu} \end{pmatrix}\,, 
  \qquad
  \bar{\Theta}_{\mu}{}^{M} = \bar{e}_{0\mu}{}^{\bar{m}} \big(\bar{V}_{0}^{t}\big)_{\bar{m}}{}^{M} = \frac{1}{\sqrt{2}} \begin{pmatrix} \delta_{\mu}{}^{\nu} \\ -\eta_{\mu\nu}\end{pmatrix}\,.
\label{Thetas}\end{equation}
We may convert the $\mathit{O}(D,D)$ vector indices, $M,N,P,\cdots$, into the conventional $D$-dimensional vector indices, $\mu,\nu,\rho,\cdots$, using them.

From the properties of $V_{0}$ and $\bar{V}_{0}$, $\Theta_{\mu}{}^{M}$ and $\bar{\Theta}_{\mu}{}^{M}$ satisfy the following relations
\begin{equation}
\begin{aligned}
  &P_{0}^{MN} = \eta^{\mu\nu} \Theta_{\mu}{}^{M} \Theta_{\nu}{}^{N}\,, \qquad \bar{P}_{0}^{MN} = -\eta^{\mu\nu} \bar{\Theta}_{\mu}{}^{M} \bar{\Theta}_{\nu}{}^{N}\,,
  \\
  &\Theta_{\mu}{}^{M} \Theta_{\nu}{}^{N}\mathcal{J}_{MN} = \eta_{\mu\nu}\,, \qquad \bar{\Theta}_{\mu}{}^{M} \bar{\Theta}_{\nu}{}^{N}\mathcal{J}_{MN} = - \eta_{\mu\nu}\,,\qquad \Theta_{\mu}^{M} \bar{\Theta}_{\nu}{}^{N} \mathcal{J}_{MN} = 0
\end{aligned}\label{}
\end{equation}
Using these operators, we denote the projected perturbations of $\mathcal{H}$ as 
\begin{equation}
\begin{aligned}
  \mathcal{E}_{\mu\nu} &= \Theta_{\mu}{}^{M}\bar{\Theta}_{\nu}{}^{N} \mathcal{E}_{MN}= \Theta_{\mu}{}^{M}\bar{\Theta}_{\nu}{}^{N} \Pi_{MN}\,, &\qquad \bar{\mathcal{E}}_{\mu\nu} &= \bar{\Theta}_{\mu}{}^{M}\Theta_{\nu}{}^{N} \bar{\mathcal{E}}_{MN}= \bar{\Theta}_{\mu}{}^{M}\Theta_{\nu}{}^{N} \Pi_{MN}\,, 
  \\
  \Delta_{\mu\nu} &= \Theta_{\mu}{}^{M}\Theta_{\nu}{}^{N} \Delta_{MN}= \Theta_{\mu}{}^{M}\Theta_{\nu}{}^{N} \Pi_{MN}\,, &\qquad \bar{\Delta}_{\mu\nu} &= \bar{\Theta}_{\mu}{}^{M} \bar{\Theta}_{\nu}{}^{N} \bar{\Delta}_{MN}= \bar{\Theta}_{\mu}{}^{M} \bar{\Theta}_{\nu}{}^{N} \Pi_{MN}\,.
\end{aligned}\label{fluctuation_D}
\end{equation}

Let us confine ourselves to the pure gravity and ignore other massless NSNS fields, $B$ and $\phi$. Then the generalised metric and the DFT dilaton are simply 
\begin{equation}
  \mathcal{H}_{MN} = \begin{pmatrix} g^{\mu\nu} & 0 \\ 0 & g_{\mu\nu} \end{pmatrix} \,, \qquad e^{-2d} = \sqrt{-g}\,,
\label{para_pureGR}\end{equation}
and we have the following additional relations among the fluctuations
\begin{equation}
  \mathcal{E}_{\mu\nu} = \bar{\mathcal{E}}_{\mu\nu}\,, \qquad \Delta_{\mu\nu} = \bar{\Delta}_{\mu\nu}\,.
\label{}\end{equation}
Then the expansion of the $\Delta_{\mu\nu}$ (or $\bar{\Delta}_{\mu\nu}$) in terms of $\mathcal{E}_{\mu\nu}$ \eqref{D_Db_expansion} reduces to
\begin{equation}
\begin{aligned}
  \Delta_{\mu\nu} = \bar{\Delta}_{\mu\nu} &= -\eta_{\mu\rho}\Big(\mathbf{1}-\sqrt{\mathbf{1}+\eta^{-1}\mathcal{E}\eta^{-1}\mathcal{E}}\Big){}^{\rho}{}_{\nu}
  \\
  &= \mathcal{E}_{\mu\rho} \Big(\frac{1}{2} \mathcal{E}^{\rho}{}_{\nu} - \frac{1}{8} (\mathcal{E}^{3})^{\rho}{}_{\nu} + \frac{1}{16} (\mathcal{E}^{5})^{\rho}{}_{\nu} + \cdots\Big)                                                                                                                                                                                                                                                                                                                 
\end{aligned}\label{Delta_in_calE}
\end{equation}

We now compare $\mathcal{E}_{\mu\nu}$ and $\Delta_{\mu\nu}$ with the metric perturbation, $g_{\mu\nu} = \eta + h_{\mu\nu}$ or $g^{\mu\nu} = \eta^{\mu\nu} - \tilde{h}^{\mu\nu}$, where $\tilde{h}^{\mu\nu} = h^{\mu\nu} -(h^{2})^{\mu\nu} + (h^{3})^{\mu\nu} -\cdots$. Using the parametrisation of the generalised metric \eqref{para_H}, we can read off what is $\Pi_{MN}$ in terms of the metric fluctuations when $B=0$,
\begin{equation}
  \Pi_{MN} = \begin{pmatrix} - h^{\mu\nu} & 0 \\ 0 & \tilde{h}_{\mu\nu} \end{pmatrix}\,.
\label{}\end{equation}
Substituting the $\Pi_{MN}$ into \eqref{fluctuation_D}, we can represent $\mathcal{E}$ and $\Delta$ in terms of $h$ and $\tilde{h}$
\begin{equation}
\begin{aligned}
  \mathcal{E}_{\mu\nu} = \frac{1}{2} \big(h + \tilde{h}\big)_{\mu\nu}\,, \qquad \Delta_{\mu\nu} =  \frac{1}{2} \big(-h + \tilde{h}\big)_{\mu\nu}\,,
\end{aligned}\label{ED_hht}
\end{equation}
or
\begin{equation}
  g^{\mu\nu} = \eta^{\mu\nu} -\mathcal{E}^{\mu\nu} + \Delta^{\mu\nu}\,, \qquad g_{\mu\nu} = \eta_{\mu\nu } + \mathcal{E}_{\mu\nu} + \Delta_{\mu\nu}
\label{ED_hht2}\end{equation}
This shows the equivalence of perturbative expansion of the generalised metric and the usual metric perturbation in GR up to field redefinition. Furthermore, it is interesting to note that the relation between $\mathcal{E}$ and $\Delta$ is not restricted to the linearised perturbation. Therefore, $\mathcal{E}$ can encodes not even linear perturbation $h_{\mu\nu}$, but also all the higher order perturbation of metric $h^{{\scriptscriptstyle(n)}}$ with $n>1$ by the following expansion
\begin{equation}
  \kappa \mathcal{E}_{\mu\nu} = \kappa h_{\mu\nu} +\sum_{n=2}^{\infty} k^{n} h^{{\scriptscriptstyle(n)}}_{\mu\nu}.
\label{}\end{equation}
%

\subsection{Perturbation of DFT equations of motion for pure gravity}
We now discuss the perturbative expansion of the equations of motion (EoM) of DFT. Similar to GR, the generalised curvature tensors are the EoMs of DFT. The EoM of the DFT dilaton and the generalised metric are the generalised curvature scalar $\mathcal{R}$ and the generalised curvature tensor $\mathcal{R}_{MN}$ respectively \cite{Jeon:2010rw,Jeon:2011cn,Hohm:2010pp}. Note that the generalised curvature scalar is the DFT Lagrangian
\begin{equation}
\begin{aligned}
  \mathcal{R} &= 4 \mathcal{H}^{M N} \partial_{M} \partial_{N} d-\partial_{M} \partial_{N} \mathcal{H}^{M N} 
   -4 \mathcal{H}^{M N} \partial_{M} d \partial_{N} d
   +4 \partial_{M} \mathcal{H}^{M N} \partial_{N} d 
   \\
   &\quad 
   +\frac{1}{8} \mathcal{H}^{M N} \partial_{M} \mathcal{H}^{K L} \partial_{N} \mathcal{H}_{K L}-\frac{1}{2} \mathcal{H}^{M N} \partial_{M} \mathcal{H}^{K L} \partial_{K} \mathcal{H}_{N L} \,,
\end{aligned}\label{}
\end{equation}
Since the variation of the generalised metric has to be constrained by the $\mathit{O}(D,D)$ constraint \eqref{ODD_const}, $\delta \mathcal{H}_{MN} = P_{(M}{}^{P}\bar{P}_{N)}{}^{Q}\delta \mathcal{H}_{PQ}$, EoM of $\mathcal{H}$ not an arbitrary variation of the Lagrangian with respect to $\mathcal{H}$, but we have to remove the unphysical variations in $\delta \mathcal{H}$ using the projectors
\begin{equation}
\begin{aligned}
  \mathcal{R}_{KL} &= P_{K}{}^{M}\bar{P}_{L}{}^{N} \mathcal{K}_{MN}\,,
\end{aligned}\label{}
\end{equation}
where $\mathcal{K}_{MN} =  \frac{\delta \mathcal{L}}{\delta \mathcal{H}^{MN}}$,
\begin{equation}
\begin{aligned} 
  \mathcal{K}_{M N} = &\ \frac{1}{8} \partial_{M} \mathcal{H}^{PQ} \partial_{N} \mathcal{H}_{PQ}-\frac{1}{4} \big(\partial_{P}-2 \partial_{P} d\big) \Big(\mathcal{H}^{PQ} \partial_{Q} \mathcal{H}_{M N}\Big) +2 \partial_{M} \partial_{N} d 
  \\ 
  &-\frac{1}{2} \partial_{(M} \mathcal{H}^{PQ} \partial_{|P|} \mathcal{H}_{N) Q}+\frac{1}{2}\big(\partial_{P}-2\partial_{P} d\big)\Big(\mathcal{H}^{PQ} \partial_{(M} \mathcal{H}_{N) Q}+\mathcal{H}_{(M}{}^{Q} \partial_{Q} \mathcal{H}_{N)}{}^{P}\Big)\,.
\end{aligned}\label{}
\end{equation}

The gauge symmetry of DFT is the generalised diffeomorphism or the generalised Lie derivative which acts on the DFT fields as
\begin{equation}
\begin{aligned}
  \hat{\mathcal{L}}_{X} \mathcal{H}_{M N} &= X^{P} \partial_{P} \mathcal{H}_{M N}+\left(\partial_{M} X^{P}-\partial^{P} X_{M}\right) \mathcal{H}_{P N}+\left(\partial_{N} X^{P}-\partial^{P} X_{N}\right) \mathcal{H}_{M P} \,,
  \\
  \hat{\mathcal{L}}_{X} d &= X^{M} \partial_{M} d-\frac{1}{2} \partial_{M} X^{M} \,,
\end{aligned}\label{gen_Lie_der}
\end{equation}
where the generalised metric $\mathcal{H}$ is a rank-2 tensor and the DFT dilaton $d$ is a scalar density with respect to the generalised Lie derivative. The gauge parameter $X^{M}$ combines the diffeomorphism parameter $\xi^{\mu}$ and the one-form gauge parameter $\Lambda_{\nu}$ for the Kalb-Ramond field in an $O(D,D)$ covariant manner
\begin{equation}
  X_{M}= \begin{pmatrix} \xi^{\mu} \\ \Lambda_{\mu}\end{pmatrix}\,.
\label{gen_Lie_para}\end{equation}
Then the infinitesimal transformation of $\mathcal{E}_{MN}$ and $\Delta_{MN}$ are
\begin{equation}
\begin{aligned}
  \hat{\mathcal{L}}_{X} \mathcal{E}_{MN} &= 2 P_{0M}{}^{P} \bar{P}_{0N}{}^{Q} \Big(\partial_{Q} X_{P} - \partial_{P} X_{Q}\Big) \,,
  \\
  \hat{\mathcal{L}}_{X} \Delta_{MN} &= 0\,.
\end{aligned}
\end{equation}

To remove the gauge degrees of freedom, we have to impose a proper gauge fixing condition. Here we introduce the following gauge condition 
\begin{equation}
  \partial_{M}\big(e^{-2d} \mathcal{H}^{MN}\big) = 0\,.
\label{gauge_condition}\end{equation}
We may rewrite this gauge condition in undoubled $D$-dimensional form by using the projectors, $\Theta_{\mu}{}^{M}$ and $\bar{\Theta}_{\mu}{}^{M}$, and substituting the perturbation of $\mathcal{H}_{MN}$ \eqref{fluctuation_D}
\begin{equation}
  \partial^{\mu} \big(\mathcal{E}_{\mu\nu}-\Delta_{\mu\nu}\big) +2 \partial^{\mu} d \big(\eta_{\mu\nu} -\mathcal{E}_{\mu\nu} +\Delta_{\mu\nu}\big) =0 \,.
\label{gauge_condition_d}\end{equation}
Obviously it is a nonlinear gauge condition in $\mathcal{E}_{\mu\nu}$ because $\Delta_{\mu\nu}$ is expanded by $\mathcal{E}_{\mu\nu}$ \eqref{Delta_in_calE}. One can show that this is the same as the de Donder gauge condition in GR (or harmonic coordinate condition) from \eqref{DFT_dilaton} and \eqref{ED_hht2}
\begin{equation}
  \partial_{\mu} \big(\sqrt{-g} g^{\mu\nu}\big)= 0 \,.
\label{}\end{equation}
Since the DFT EoM for pure gravity is equivalent to the vacuum Einstein equation, our gauge condition \eqref{DFT_eom_gauge_condition} also provides a well-posed initial value problem of Cauchy. However, great care needs to be taken with consistency of the gauge choice in the case where the dilaton is nontrivial \cite{Choquet-Bruhat,Choquet-Bruhat:1985xei}. If we introduce the linearised metric perturbation $\sqrt{-g} g^{\mu\nu} = \eta^{\mu\nu} + \mathfrak{h}^{\mu\nu}$\footnote{It is the opposite convention to the usual metric perturbation $\sqrt{-g} g_{\mu\nu} = \eta_{\mu\nu} + \mathfrak{h}_{\mu\nu}$. These two choices are related by a field redefinition.}, then the de Donder gauge condition reduces to
\begin{equation}
  \partial_{\mu} \mathfrak{h}^{\mu\nu} = 0,
\label{inverse_de_Donder}\end{equation}
and it is equivalent to \eqref{gauge_condition_d}.

Imposing the gauge fixing condition \eqref{gauge_condition}, the equations of motion reduces to
\begin{equation}
\begin{aligned}
  \mathcal{R} &= 2\mathcal{H}^{M N} \partial_{M} \partial_{N} d
   +\frac{1}{8} \mathcal{H}^{M N} \partial_{M} \mathcal{H}^{K L} \partial_{N} \mathcal{H}_{K L}-\frac{1}{2} \mathcal{H}^{M N} \partial_{M} \mathcal{H}^{K L} \partial_{K} \mathcal{H}_{N L} \,,
  \\
  \mathcal{R}_{KL} &=  P_{K}{}^{M}\bar{P}_{L}{}^{N} \tilde{\mathcal{K}}_{MN}\,,
\end{aligned}\label{DFT_eom_gauge_condition}
\end{equation}
where $\tilde{\mathcal{K}}_{MN}$ is the gauge fixed version of $\mathcal{K}_{MN}$
\begin{equation}
\begin{aligned}
  \tilde{\mathcal{K}}_{MN} &= \frac{1}{2} \partial_{M} P^{PQ} \partial_{N}P_{PQ} %
  - P^{PQ} \partial_{P} \partial_{Q} P_{MN} %
  -4 \partial_{(M} P^{PQ} \partial_{|P|} P_{N)Q} %
  +2\partial^{P}P_{MQ} \partial^{Q}P_{NP} \,.
\end{aligned}\label{}
\end{equation}
Interestingly, DFT dilaton $d$ does not appear in $\mathcal{R}_{MN}$ after the gauge fixing. This is no surprise, because for pure gravity $d$ does not give a new degree of freedom - it is just the determinant of metric.

Note that the components of $\mathcal{R}_{MN}$ are not independent - it contains redundant equations due to the manifest $\mathit{O}(D,D)$ invariance. There are total $(2D)^{2}$ components in $\mathcal{R}_{MN}$, but the number of independent equations should be $D^{2}$ only: $\frac{D(D+1)}{2}$ for $g_{\mu\nu}$ and $\frac{D(D-1)}{2}$ for $B_{\mu\nu}$. To remove the redundancies, we project out $\mathcal{R}_{MN}$ to the $D$-dimensional undoubled form using $\Theta_{\mu}{}^{M}$ and $\bar{\Theta}_{\mu}{}^{M}$ \eqref{Thetas}. There are four distinct equations,
\begin{equation}
\begin{aligned}
  \Theta_{\mu}{}^{M} \Theta_{\nu}{}^{N} \mathcal{R}_{MN} : &\ -\frac{1}{2} \Theta_{\mu}{}^{M} \Theta_{\nu}{}^{N} \Big(\Pi_{N}{}^{P} \tilde{\mathcal{K}}_{MP} %
  -\frac{1}{2} \Pi_{M}{}^{P} \Pi_{N}{}^{Q} \tilde{\mathcal{K}}_{PQ}\Big) =0 \,, %
  \\
  \Theta_{\mu}{}^{M}\bar{\Theta}_{\nu}{}^{N} \mathcal{R}_{MN} : &\ \Theta_{\mu}{}^{M}\bar{\Theta}_{\nu}{}^{N} \Big(\tilde{\mathcal{K}}_{MN} %
  -\frac{1}{2} \Pi_{N}{}^{P} \tilde{\mathcal{K}}_{MP} %
  +\frac{1}{2} \Pi_{M}{}^{P} \tilde{\mathcal{K}}_{PN} %
  -\frac{1}{4} \Pi_{M}{}^{P} \Pi_{N}{}^{Q} \tilde{\mathcal{K}}_{PQ}\Big) =0 \,, %
  \\
  \bar{\Theta}_{\mu}{}^{M}\Theta_{\nu}{}^{N} \mathcal{R}_{MN} : &\ -\frac{1}{4} \bar{\Theta}_{\mu}{}^{M}\Theta_{\nu}{}^{N} \Pi_{M}{}^{P} \Pi_{N}{}^{Q} \tilde{\mathcal{K}}_{PQ} =0 \,,
  \\
  \bar{\Theta}_{\mu}{}^{M}\bar{\Theta}_{\nu}{}^{N} \mathcal{R}_{MN} : &\  \frac{1}{2} \bar{\Theta}_{\mu}{}^{M}\bar{\Theta}_{\nu}{}^{N}\Big(\Pi_{M}{}^{P} \tilde{\mathcal{K}}_{PN} %
  -\frac{1}{2} \Pi_{M}{}^{P} \Pi_{N}{}^{Q} \tilde{\mathcal{K}}_{PQ}\Big) =0 \,.
\end{aligned}\label{decomp_eom}
\end{equation}
As one can see, the chiral and the antichiral parts, $\Theta_{\mu}{}^{M} \Theta_{\nu}{}^{N} \mathcal{R}_{MN} $ and $\bar{\Theta}_{\mu}{}^{M}\bar{\Theta}_{\nu}{}^{N} \mathcal{R}_{MN}$, can be represented by the mixed chirality parts, $\Theta_{\mu}{}^{M}\bar{\Theta}_{\nu}{}^{N} \mathcal{R}_{MN}$ and $\bar{\Theta}_{\mu}{}^{M} \Theta_{\nu}{}^{N} \mathcal{R}_{MN}$. Finally, combining the two mixed chirality sectors, we have the minimal EoM as
\begin{equation}
\begin{aligned}
  &\Theta_{\mu}{}^{M}\bar{\Theta}_{\nu}{}^{N} \Big(\tilde{\mathcal{K}}_{MN} %
  -\frac{1}{2} \tilde{\mathcal{K}}_{MP} \mathcal{E}^{P}{}_{N} %
  -\frac{1}{2} \tilde{\mathcal{K}}_{MP} \bar{\Delta}^{P}{}_{N} %
  +\frac{1}{2} \mathcal{E}_{M}{}^{P} \tilde{\mathcal{K}}_{PN}%
  +\frac{1}{2} \Delta_{M}{}^{P} \tilde{\mathcal{K}}_{PN}\Big) =0\,.
\end{aligned}\label{}
\end{equation}
We now present the perturbative DFT equations of motion in undoubled $D$-dimensional form using $\Theta_{\mu}{}^{M}$ and $\bar{\Theta}_{\mu}{}^{M}$
\begin{tcolorbox}
\begin{equation}
\begin{aligned}
  \big(\delta+\Delta\big){}_{(\mu}{}^{\rho}\big(X+Y+Z-\Box\mathcal{E}\big)}{_{\nu)\rho} - \mathcal{E}_{(\mu}{}^{\rho}\big(X -Y -W\big){}_{\nu)\rho} =0\,,
\end{aligned}\label{eom}
\end{equation}
where $X_{\mu\nu}$, $Y_{\mu\nu}$, $Z_{\mu\nu}$ and $W_{\mu\nu}$ are auxiliary fields defined as
\begin{equation}
\begin{aligned}
  X_{\mu\nu} &= -\frac{1}{2} \partial_{(\mu}(\mathcal{E}-\Delta )^{\rho\sigma} \partial_{\nu)} (\mathcal{E}+\Delta )_{\rho\sigma} + 2\partial_{(\mu}(\mathcal{E}-\Delta )^{\rho\sigma}\partial_{|\rho|}(\mathcal{E} +\Delta )_{\nu)\sigma}\,,
  \\
  Y_{\mu\nu} &=-\partial^{\rho} (\mathcal{E}-\Delta )_{\mu \sigma} \partial^{\sigma} (\mathcal{E}-\Delta )_{\nu \rho}\,,
  \\
  Z_{\mu\nu} &= (\mathcal{E}-\Delta)^{\rho\sigma}\partial_{\rho}\partial_{\sigma} \mathcal{E}_{\mu\nu}\,, \qquad W_{\mu\nu} = \big(\eta-\mathcal{E}+\Delta \big)^{\rho\sigma}\partial_{\rho}\partial_{\sigma} \Delta_{\mu\nu} \,.
\end{aligned}\label{DFT_perturbative_eom}
\end{equation}
\end{tcolorbox}
This equation describes the graviton dynamics in terms of the generalised metric perturbations, $\mathcal{E}_{\mu\nu}$ and $\Delta_{\mu\nu}$, which yields the same result as perturbative Einstein equation for the pure gravity. 

\section{Off-Shell Recursion for Gravity}\label{Sec:3}
Berends-Giele (BG) recursion in Yang-Mills theory provides an efficient algorithm for computing color-ordered amplitudes iteratively \cite{Berends:1987me}. The recursion was obtained by using the recursive structure of YM's Feynman vertices. The main ingredient of this technique is the off-shell current $J_{\mu}^{12\cdots n}$ corresponding to a Feynman diagram with $n$ on-shell legs and one off-shell leg. The off-shell currents give $(n+1)$-point color ordered amplitude $A(1,2,\cdots,n+1)$ as follows\footnote{Throughout this paper, we will not distinguish the position of the spacetime indices in the off-shell currents.}:
\begin{equation}
  A(1,2,\cdots,n+1) = \lim_{s_{12\cdots n}\to 0} s_{12\cdots n} \epsilon^{n+1}_{\mu} J_{\mu}^{12\cdots n}
\label{amplitude_current_gluon}\end{equation}
where $\epsilon^{n+1}_{\mu}$ is the polarization vector for the $(n+1)$-th external gluon and $s_{12\cdots n} = - (k^{1}+k^{2}+\cdots + k^{n})^{2}$. Note that only simple poles with respect to $s^{12\cdots n}$ of the currents contribute to the amplitudes. 

However, the conventional method deriving the off-shell recursion relations using Feynman vertices is not available for perturbative gravity due to the infinitely many irregular Feynman vertices. Unlike the YM case, new Feynman vertices arise as the number of external legs is increased. To circumvent this problem, we will adopt another approach, the so-called perturbiner expansion, which uses equations of motion rather than Feynman vertices. Recently, the perturbiner method has been applied to the perturbative GR \cite{Gomez:2021shh}. Here, we will construct the graviton recursion relation using the perturbiner method for the perturbative DFT obtained in the previous section.

\subsection{Perturbiner method for Yang-Mills theory}\label{Sec:3.1}
We now briefly review the perturbiner expansion, which generates the BG currents and their recursion relations. The original formulation of the BG recursion relation due to Berends and Giele considers the structure of Feynman diagrams by defining the off-shell currents, which is a Feynman diagram with one off-shell leg. On the other hand, the perturbiner expansion is a multi-particle expansion of the Lie algebra valued gauge field $\mathbb{A}_{\mu}$ in the plane-wave basis \cite{Rosly:1996vr,Rosly:1997ap,Selivanov:1997aq,Selivanov:1997an,Selivanov:1997ts}
\begin{equation}
  \mathbb{A}_{\mu} = \sum_{i} J_{\mu}^{i} T^{a_{i}} e^{ik^{i}\cdot x} 
  +\sum_{i, j} J^{i j}_{\mu} T^{a_{i}} T^{a_{j}} e^{ik^{i j} \cdot x} 
  +\sum_{i,j,k} J^{ijk}_{\mu} T^{a_{i}} T^{a_{j}}T^{a_{k}} e^{ik^{ijk} \cdot x}+\cdots \,. 
\label{perturbiner_A}
\end{equation}
where the coefficients of the expansion, $J_{\mu}^{i}$, $J_{\mu}^{ij},\cdots$, are the BG currents \cite{Mafra:2015gia,Lee:2015upy,Mafra:2015vca} and $i,j,k\cdots$ are \textit{letters} which represent single-particle labels. Here $T^{a_{i}}$ are the Lie group generators, and $k^{ijk\cdots}_{\mu} = k^{i}_{\mu} + k^{j}_{\mu}+ k^{k}_{\mu}+ \cdots$. We may simplify the expansion as
\begin{equation}
  \mathbb{A}_{\mu} = \sum_{P} J_{\mu}^{P} T^{P}e^{ik_{P}\cdot x}
\label{gluon_perturbiner_expansion}\end{equation}
where 
\begin{equation}
  T^{P} = T^{a_{i}} T^{a_{j}}T^{a_{k}} \cdots \,, \qquad k^{P}_{\mu} = k^{i}_{\mu} + k^{j}_{\mu}+ k^{k}_{\mu}+ \cdots .
\label{}\end{equation}
and $P,Q,R\cdots$ are \textit{words} which consist of the letters, such as $P=ijkl\cdots $, which are multi-particle labels. We call the length of the words `rank' and denote as $|P|$, $|Q|$ and $|R|$ etc. 

The conventional method of constructing the BG recursion relations is using the recursive structure of Feynman diagrams. On the other hand, in the perturbiner method, the BG recursion arises by substituting the perturbiner expansion into the EoM. Thus the perturbiner expansions are solutions of the EoM. Since scattering amplitudes are associated with the BG currents, the perturbiner method connects the tree-level scattering amplitudes and solutions of equations of motion manifestly. 


The EoM of the YM theory in the Lorentz gauge, $\partial^{\mu} \mathbb{A}_{\mu} = 0$, is given by 
\begin{equation}
\begin{aligned}
  \Box \mathbb{A}_{\mu} &= \frac{i}{\sqrt{2}} [ \mathbb{A}^{\nu} , \partial_{\nu} \mathbb{A}_{\mu} + \mathbb{F}_{\nu \mu}]\,,
  \\
  \mathbb{F}_{\mu\nu} &=  \partial_{\mu} \mathbb{A}_{\nu} - \partial_{\nu} \mathbb{A}_{\mu} - \frac{i}{\sqrt{2}} [ \mathbb{A}_{\mu} , \mathbb{A}_{\nu} ]\,,
\end{aligned}\label{}
\end{equation}
where our Lie algebra convention is
\begin{equation}
  \text{Tr}(T^{a} T^{b}) = \delta^{ab} \,, \qquad  [T^{a},T^{b}] = i f^{abc}T^{c}\,.
\label{}\end{equation}
Note that we treat the field strength $\mathbb{F}_{\mu\nu}$ as an auxiliary field to avoid the double commutators in the EoM. Then it requires a perturbiner expansion associated $\mathbb{F}_{\mu\nu}$ 
\begin{equation}
\begin{aligned}
  \mathbb{F}_{\mu \nu} = \sum_{P} F_{\mu \nu}^{P} T^{P}e^{ i k_{P} \cdot x}\,.
\end{aligned}\label{}
\end{equation}

As we have mentioned, the gluon recursion can be obtained by substituting the perturbiner expansion \eqref{perturbiner_A} into the above EoM 
\begin{equation}
\begin{aligned}
  J^{P}_{\mu} &=  \frac{i}{\sqrt{2} s^{P}}\sum_{ P = QR} \Big( i ( J^{Q} \cdot k^{R} ) J^{R}_{\mu} + J^{Q}_{\nu} F^{R}_{\nu \mu} - ( Q \leftrightarrow R ) \Big)\,,
  \\
  F^{P}_{\mu \nu} &= i k^{P}_{\mu} J^{P}_{\nu} - i k^{P}_{\nu} J^{P}_{\mu} - \frac{i}{\sqrt{2}} \sum_{ P = QR} \Big( J^{Q}_{\mu} J^{R}_{\nu} -  ( Q \leftrightarrow R ) \Big)\,,
\end{aligned}\label{YM_recursions}
\end{equation}
where the sum goes over all deconcatenations of the given word $P = p_{1} p_{2} \cdots p_{|P|} $ into two words $Q=p_{1} p_{2} \cdots p_{j}$ and $R=p_{j+1} p_{j+2} \cdots p_{|P|}$ for $1\leq j \leq |P|$. For instance, for a few word $P$, all possible $(Q,R)$ pairs are
\begin{equation}
\begin{aligned}
  P &= 12 &\quad (Q,R) &= (1,2) \,,
  \\
  P &= 123 &\quad (Q,R) &= (1,23),\ (12,3) \,,
  \\
  P &= 1234 &\quad (Q,R) &= (1,234),\ (12,34),\ (123,4) \,.
\end{aligned}\label{}
\end{equation}
The initial condition of the BG recursion is identified with the polarization vector, $J_{\mu}^{i} = \epsilon^{i}_{\mu}$. We list the explicit expression of the gluon off-shell currents in appendix \ref{appendix_B}

\subsection{Perturbiner expansion for DFT} \label{sec:3.2}

We now define the perturbiner expansion for graviton and construct the off-shell recursion relations for pure gravity. The graviton perturbiner expansion is not the same as the gluon case. The main difference in the graviton currents is the colour indices. For gluon currents, matrix products of the Lie algebra generators $T^{i}T^{j}T^{k}\cdots$ are used to organize the perturbiner expansion, and the ordering of letters is important due to the noncommutativity of $T^{i}$. However, graviton field $\mathcal{E}_{\mu\nu}$ does not carry colour indices, and we can only use the plane-wave basis $e^{i k_{P}{\cdot}x}$ to separate the states in the expansion \cite{Mafra:2016ltu,Mizera:2018jbh}. Since the letters in the plane-wave basis are commute, the ordering of letters in graviton perturbiners is irrelevant. Thus the graviton currents satisfy $\mathcal{J}_{\mu\nu}^{\alpha} = \mathcal{J}_{\mu\nu}^{\beta}$ for any $\alpha,\beta\in  S_{n}$, where $S_{12\cdots n}$ is a permutation group with the set $\{1,2,\cdots,n\}$. 

We introduce the graviton perturbiner expansion using the ordered words 
\begin{equation}
\begin{aligned}
  \mathcal{E}_{\mu\nu} &= 
    \sum_{i} \mathcal{J}_{\mu\nu}^{i} e^{i k_{i}\cdot x} 
  + \sum_{i<j} \mathcal{J}_{\mu\nu}^{ij} e^{i k_{ij}\cdot x} 
  + \sum_{i<j<k} \mathcal{J}_{\mu\nu}^{ijk} e^{i k_{ijk}\cdot x} + \cdots\,,
  \\
  &= \sum_{\mathcal{P}} \mathcal{J}_{\mu\nu}^{\mathcal{P}} e^{i k_{\mathcal{P}}\cdot x}\,,
\end{aligned}\label{perturbiner_expansion_E}
\end{equation}
where $\mathcal{J}_{\mu\nu}^{\mathcal{P}}$ is the graviton off-shell currents and $\mathcal{P}$ is an ordered words $\mathcal{P} = p_{1}p_{2}\cdots p_{|\mathcal{P}|}$ with $p_{1} < p_{2} <\cdots < p_{|\mathcal{P}|}$. Comparing with the words for gluons, for instance a length 2 word for gluon $P=12$ is different from $21$, however, a length 2 words for graviton $\mathcal{P}=12 = 21$. Like the gluon perturbiner expansion, the coefficient of the expansion $\mathcal{J}_{\mu\nu}^{\mathcal{P}}$ are the graviton BG currents. Since the EoM of the perturbative DFT \eqref{DFT_perturbative_eom} consists of the additional auxiliary fields $\Delta_{\mu\nu}$, $X_{\mu\nu}$, $Y_{\mu\nu}$, $Z_{\mu\nu}$ and $W_{\mu\nu}$ other than $\mathcal{E}_{\mu\nu}$, we introduce their perturbiner expansions as well
\begin{equation}
\begin{aligned}
  \Delta_{\mu\nu} &= \sum_{\mathcal{P}} \varDelta_{\mu\nu}^{\mathcal{P}} e^{i k_{\mathcal{P}}\cdot x}
  \,,
  \qquad
  X_{\mu\nu} = \sum_{\mathcal{P}} \mathcal{X}_{\mu\nu}^{\mathcal{P}} e^{i k_{\mathcal{P}}\cdot x}\,,
  \qquad
  Y_{\mu\nu} = \sum_{\mathcal{P}} \mathcal{Y}_{\mu\nu}^{\mathcal{P}} e^{i k_{\mathcal{P}}\cdot x}\,,
  \\
  Z_{\mu\nu} &= \sum_{\mathcal{P}} \mathcal{Z}_{\mu\nu}^{\mathcal{P}} e^{i k_{\mathcal{P}}\cdot x}\,,
  \qquad
  W_{\mu\nu} = \sum_{\mathcal{P}} \mathcal{W}_{\mu\nu}^{\mathcal{P}} e^{i k_{\mathcal{P}}\cdot x}\,.
\end{aligned}\label{perturbiner_expansion_aux}
\end{equation}
Note that all these auxiliary currents starts from the rank-2, \textit{i.e.} $\Delta^{i}_{\mu\nu} = \mathcal{X}^{i}_{\mu\nu} = \mathcal{Y}^{i}_{\mu\nu} = \mathcal{Z}^{i}_{\mu\nu} = \mathcal{W}^{i}_{\mu\nu} = 0$.

Recall that the BG recursion relation for YM theory can be obtained by substituting the perturbiner expansion into the YM EoM. Similarly, if we substitute the perturbiner expansions of graviton and auxiliary fields into the EoM of perturbative DFT\eqref{eom}, we have 
\begin{equation}
\begin{aligned}
  \mathcal{J}^{\mathcal{P}}_{\mu\nu} &= \frac{1}{s_{\mathcal{P}}} \bigg[\ 
    \big(\mathcal{X}^{\mathcal{P}} +\mathcal{Y}^{\mathcal{P}} +\mathcal{Z}^{\mathcal{P}}\big)_{\mu\nu} 
  + \sum_{\mathcal{P}=\mathcal{Q}\cup \mathcal{R}} 
  \mathcal{J}^{\mathcal{Q}}_{\rho(\mu} \big(- \mathcal{X}^{\mathcal{R}} +\mathcal{Y}^{\mathcal{R}} +\mathcal{W}^{\mathcal{R}}\big)_{\nu)\rho}
  \\
  &\qquad\quad + \sum_{\mathcal{P}=\mathcal{Q}\cup \mathcal{R}} \varDelta^{\mathcal{Q}}_{\rho(\mu} \big(\mathcal{X}^{\mathcal{R}}+\mathcal{Y}^{\mathcal{R}} +\mathcal{Z}^{\mathcal{R}} - s_{\mathcal{R}} \varDelta^{\mathcal{R}} \big)_{\nu)\rho}\ \bigg] \,,
\end{aligned}\label{graviton_recursion_relation}
\end{equation}
where each auxiliary current is 
\begin{equation}
\begin{aligned}
  \varDelta^{\mathcal{P}}_{\mu\nu} &= \frac{1}{2} (\mathcal{J}^{2})^{\mathcal{P}}_{\mu\nu}
  -\frac{1}{8}  \sum_{\mathcal{P}=\mathcal{Q}\cup\mathcal{R}} (\mathcal{J}^{2})^{\mathcal{Q}}_{\mu\rho} (\mathcal{J}^{2})^{\mathcal{R}}_{\nu\rho} 
  + \frac{1}{16} \sum_{\mathcal{P}=\mathcal{Q}\cup\mathcal{R}\cup\mathcal{S}} (\mathcal{J}^{2})^{\mathcal{Q}}_{\mu\rho} (\mathcal{J}^{2})^{\mathcal{R}}_{\rho\sigma}(\mathcal{J}^{2})^{\mathcal{S}}_{\nu\sigma} - \cdots\,,
  \\
  \mathcal{X}_{\mu\nu}^{\mathcal{P}} &= \sum_{\mathcal{P}=\mathcal{Q}\cup\mathcal{R}} \bigg[\,
  \frac{1}{2} k^{\mathcal{Q}}_{(\mu} k^{\mathcal{R}}_{\nu)} 
  \big(\mathcal{J}^{\mathcal{Q}}_{\rho\sigma}-\varDelta^{\mathcal{Q}}_{\rho\sigma}\big) 
  \big(\mathcal{J}^{\mathcal{R}}_{\rho\sigma} +\varDelta^{\mathcal{R}}_{\rho\sigma}\big)
  - 2 k^{\mathcal{Q}}_{(\mu} k^{\mathcal{R}}_{|\rho|} \big(\mathcal{J}^{\mathcal{Q}}_{\rho\sigma}-\varDelta^{\mathcal{Q}}_{\rho\sigma}\big) \big(\mathcal{J}^{\mathcal{R}} {+} \varDelta^{\mathcal{R}}\big)_{\nu)\sigma} \bigg] \,,
  \\
  \mathcal{Y}_{\mu\nu}^{\mathcal{P}} &= \sum_{\mathcal{P}=\mathcal{Q}\cup\mathcal{R}} 
  \big(\mathcal{J}^{\mathcal{Q}}_{\mu\rho} -\Delta^{\mathcal{Q}}_{\mu\rho}\big) k^{\mathcal{R}}_{\rho} 
  \big(\mathcal{J}^{\mathcal{R}}_{\nu\sigma} -\Delta^{\mathcal{R}}_{\nu\sigma}\big) k^{\mathcal{Q}}_{\sigma} \,,
  \\
  \mathcal{Z}^{\mathcal{P}}_{\mu\nu} &= - \sum_{\mathcal{P}=\mathcal{Q}\cup\mathcal{R}} k^{\mathcal{R}}_{\rho} k^{\mathcal{R}}_{\sigma} 
  \big(\mathcal{J}^{\mathcal{Q}}_{\rho\sigma}-\varDelta^{\mathcal{Q}}_{\rho\sigma}\big) \mathcal{J}^{\mathcal{R}}_{\mu\nu}\,,
  \\
  \mathcal{W}^{\mathcal{P}}_{\mu\nu}&= s_{\mathcal{P}} \varDelta_{\mu\nu}^{\mathcal{P}} 
  + \sum_{\mathcal{P}=\mathcal{Q}\cup\mathcal{R}} k^{\mathcal{R}}_{\rho} k^{\mathcal{R}}_{\sigma}\big(\mathcal{J}^{\mathcal{Q}}_{\rho\sigma}-\varDelta^{\mathcal{Q}}_{\rho\sigma}\big) \varDelta^{\mathcal{R}}_{\mu\nu}\,.
\end{aligned}\label{auxiliary_recursion_relation}
\end{equation}
Remarkably, all the equations are quadratic in currents except $\varDelta_{\mu\nu}^{\mathcal{P}}$. We present the explicit form of the recursion relations up to rank-4 in the appendix \ref{appendix_A}. We will solve this recursions iteratively in the next section.
As for the gluon currents, the initial condition of the graviton recursion is 
\begin{equation}
  \mathcal{J}^{i}_{\mu\nu} = \varepsilon^{i}_{\mu\nu} \,,
\label{initial_graviton}\end{equation}
where $\varepsilon^{i}_{\mu\nu}$ is the graviton polarization tensor. Further, $(n+1)$-point graviton scattering amplitudes $M(1,2,\cdots,n+1)$ can be represented by the rank-$n$ graviton off-shell current 
\begin{equation}
  M(1,2,\cdots,n+1) = \lim_{s_{12\cdots n}\to 0} s_{12\cdots n} \varepsilon^{n+1}_{\mu\nu} \mathcal{J}_{\mu\nu}^{12\cdots n}\,.
\label{}\end{equation}
%

\section{Solving the Recursions and the Classical Double Copy for Currents}\label{Sec:4}
We now iteratively solve the off-shell recursion relations obtained in the previous section. Note that the graviton currents are constructed in \cite{Cheung:2017kzx,Gomez:2021shh} by solving the recursions based on the perturbative GR. Our result is consistent with the earlier works since it gives the same graviton scattering amplitudes. Though the recursions are irrelevant to the dimensionality, we assume 4-dimensional spacetime to employ the spinor-helicity formalism. The off-shell currents should have a proper double copy structure due to their relation with the scattering amplitude. Mizera and Skrzypek proposed the KLT relation for off-shell currents of the non-gravitational effective field theories \cite{Mizera:2018jbh}. Recently Cheung and Mangan introduced a map between gluon and graviton currents based on the color-kinematic duality \cite{Cheung:2021zvb}.

Then we will present the current KLT relation for gravity by using the explicit form of the gluon and graviton currents. In the computation of the off-shell currents, we will specify helicities of the on-shell external states. We show that the current KLT relation requires regular terms for the propagator of the off-shell leg, which does not contribute to the scattering amplitude in addition to the gauge transformation terms introduced in \cite{Mizera:2018jbh}.

\subsection{Current KLT relation}
In order to see the general structure of the graviton currents, we solve the graviton recursion up to rank-3 without restriction on helicity. We present, by explicit construction, the current KLT relation generalising the conventional KLT relation for graviton scattering amplitudes.

Let us start from the rank-2 current. If we substitute the initial condition \eqref{initial_graviton}, we obtain the rank-2 currents from the recursion in \eqref{rank2_1} and \eqref{rank2_2}
\begin{equation}
\begin{aligned}
  \mathcal{J}^{12}_{\mu\nu} &= \frac{1}{2 s_{12}} \bigg[
 -2 (\epsilon_{1} {\cdot}\epsilon_{2}) \Big( k_{1}{\cdot}\epsilon_{2}  \left( k^2_{\mu} \epsilon^1_{\nu} +k^2_{\nu} \epsilon^1_{\mu} \right)
 +k_{2}{\cdot}\epsilon_{1} \left(k^1_{\mu} \epsilon^2_{\nu} +k^1_{\nu} \epsilon^2_{\mu}\right)\Big)
 \\
 &\qquad\qquad +(\epsilon_{1}{\cdot}\epsilon_{2})^{2} \left(k^2_{\mu } k^1_{\nu} +k^1_{\mu} k^2_{\nu}\right)
 -2 \Big((k_{1}{\cdot}\epsilon_{2}) \epsilon^1_{\mu} -(k_{2}{\cdot}\epsilon_{1})\epsilon^2_{\mu}\Big) \Big((k_{1}{\cdot}\epsilon_{2}) \epsilon^1_{\nu}-(k_{2}{\cdot}\epsilon_{1}) \epsilon^2_{\nu} \Big)\bigg]\,.
\end{aligned}\label{}
\end{equation}
The rank-2 gluon current is given by
\begin{equation}
  J^{12}_{\mu} = \frac{(\epsilon_{1}{\cdot}\epsilon_{2}) \left( k^2_{\mu} -k^1_{\mu}\right) +2(k_{1}{\cdot}\epsilon_{2}) \epsilon^1_{\mu } - 2(k_{2}{\cdot}\epsilon_{1}) \epsilon^2_{\mu}}{\sqrt{2} s_{12}}\,.
\label{rank-2_graviton_current}\end{equation}
Then we can show that the double copy for the BG currents. It should be consistent with the KLT relation for the 3-pt scattering amplitude
\begin{equation}
  M_{3}(1,2,3) = \frac{1}{s_{12}} A_{3}(1,2,3)A_{3}(1,2,3)\,.
\label{}\end{equation}
Using the relation between the scattering amplitude and BG currents, we expect $\mathcal{J}^{12}_{\mu\nu}$ has to accompany at least $J^{12}_{\mu} S[2|2]_{1} J^{12}_{\nu}$. If we compare them we find the following relation 
\begin{equation}
  \mathcal{J}_{\mu\nu}^{12} = J^{12}_{\mu} S[2|2]_{1} J^{12}_{\nu} + k^{12}_{\mu}\Delta_{\nu}+ k^{12}_{\mu}\Delta_{\nu} + k^{12}_{\mu}k^{12}_{\nu}\Delta \,,
\label{rank2_KLT}\end{equation}
where
\begin{equation}
\begin{aligned}
  \Delta_{\mu} = -\frac{\epsilon_{1}{\cdot}\epsilon_{2} \Big((k_{1}{\cdot}\epsilon_{2})\epsilon^1_{\mu} +(k_{2}{\cdot}\epsilon_{1})\epsilon^2_{\mu}\Big)}{2 s_{12}}\,,
  \qquad
  \Delta = \frac{(\epsilon_{1}{\cdot}\epsilon_{2})^2}{4 s_{12}} \,.
\end{aligned}\label{}
\end{equation}
The last three terms on the righthand side of \eqref{rank2_KLT} can be interpreted to the gauge transformation of the graviton field, and it does not contribute to the scattering amplitudes because $k^{12}_{\mu}$ is orthogonal with the $\epsilon^{3}$ in the on-shell limit, $k^{12}\cdot \epsilon^{3} = 0$.

Now let us consider the rank-3 currents. Note that the KLT relation for 4-pt graviton scattering amplitude is 
\begin{equation}
  M(1,2,3,4) = \frac{1}{s_{123}} \begin{pmatrix} A(1,2,3,4)&A(1,3,2,4)\end{pmatrix} \begin{pmatrix} S[23|23]_{1} & S[23|32]_{1} \\ S[32|23]_{1} &S[32|32]_{1} \end{pmatrix} 
  \begin{pmatrix} A(1,2,3,4) \\ A(1,3,2,4) \end{pmatrix}\,,
\label{}\end{equation}
where the components of the KLT kernel are
\begin{equation}
\begin{aligned}
  S[23 | 23]_{1} &= \frac{s_{12}(s_{13}+s_{23})}{4}  \,,
  \qquad S[23 | 32]_{1} = \frac{s_{12} s_{13} }{4} \,, 
  \qquad  S[32 | 32]_{1} = \frac{s_{13}(s_{12}+s_{23})}{4} 
\end{aligned}\label{}
\end{equation}
Thus the 4-pt gluon color ordered amplitudes contribute to the KLT relation are $A(1,2,3,4)$ and $A(1,3,2,4)$.

From the relation between the BG current and color-ordered amplitude, $A(1,2,3,4)$ and $A(1,3,2,4)$ are associated with $J^{123}_{\mu}$ and $J^{132}_{\mu}$ respectively. Solving the rank-3 recursion relation \eqref{YM_recursions} gives
\begin{equation}
\begin{aligned}
  J_{\mu}^{123} &= \frac{2}{s_{123}} \Bigg[
 \bigg( \frac{ Q_{132} k^2_{\mu} - Q_{231} k^{1}_{\mu} 
  + R_{123} k^3_{\mu}}{s_{12}} + (1\leftrightarrow 3)\bigg)  \\
  &\qquad\quad - \bigg(\left(
  P_{123}+P_{132} + \frac{\epsilon_{2}{\cdot}\epsilon_{3}}{2}\right) \frac{s_{13}}{s_{23}}\epsilon^1_{\mu} +\text{cycl[1,2,3]}\bigg)\ \Bigg]
  + \text{regular terms} + k^{123}_{\mu} \Psi^{123} \,,
\end{aligned}\label{J123A}
\end{equation}
where $P_{ijk}$, $Q_{ijk}$ and $R_{ijk}$ are defined by
\begin{equation}
\begin{aligned}
  P_{ijk} &= \frac{k_{i}{\cdot}\epsilon_{j}\big(k_{i}{\cdot}\epsilon_{k} +k_{j}{\cdot}\epsilon_{k}\big)}{s_{ij}}\,,
  \qquad 
  s_{ij} P_{ijk} - s_{ik} P_{ikj} = (k_{i}{\cdot}\epsilon_{j})(k_{j}{\cdot}\epsilon_{k}) - (k_{i}{\cdot}\epsilon_{k})(k_{k}{\cdot}\epsilon_{j})\,,
  \\
  Q_{ijk} &= (k_{i}{\cdot}\epsilon_{j})(\epsilon_{i}{\cdot}\epsilon_{k})\,, 
  \qquad 
  R_{ijk} = Q_{ijk} - Q_{jik}\,.
\end{aligned}\label{PQR}
\end{equation}
Since the only simple poles in the currents contribute to the color-ordered amplitudes, we single out the regular terms in $s_{123}\to 0$
\begin{equation}
  \text{regular terms} :~  \bigg(2P_{123}+\frac{\epsilon_{2}{\cdot}\epsilon_{3}}{2}\bigg)\frac{\epsilon^1_{\mu }}{s_{23}}+\Big(2P_{321}+\frac{\epsilon_{1}{\cdot}\epsilon_{2}}{2 } \Big)\frac{\epsilon^{3}_{\mu}}{s_{12}}\,.
\label{}\end{equation}
The last term in \eqref{J123A} can be interpreted to a gauge transformation and does not contribute to the scattering amplitude
\begin{equation}
\begin{aligned}
  \Psi^{123} &= \frac{1}{2 s_{123}} \bigg[\frac{1}{s_{12}} \Big(Q_{231}-Q_{132}-2R_{123} \Big) +(1\leftrightarrow 2)\Big)\bigg]\,.
\end{aligned}\label{Psi123}
\end{equation}
Similarly, $J_{\mu}^{132}$ can be obtained by replacing $2\to 3$ and $3\to 2$. 

We now consider the rank-3 graviton current $J^{123}_{\mu}$. It is obtained by solving the recursion relation \eqref{rank3_graviton_current}, but the explicit form is rather messy. The for the double copy of the BG currents is 
\begin{equation}
\begin{aligned}
  \mathcal{J}^{123}_{\mu\nu} &= \sum_{\rho,\sigma\in S_{\{2,3\}}} J_{\mu}^{1\alpha} S[\alpha|\beta]_{1} J_{\nu}^{1\beta} +\text{regular terms}
   +2 k_{(\mu}^{123} \Delta^{123}_{\nu)} + k_{\mu}^{123} k_{\nu}^{123}\Delta^{123} \,.
\end{aligned}\label{}
\end{equation}
where $S_{\{2,3\}}$ is the permutation group of the set $\{2,3\}$, and $\Delta$ and $\Delta_{\mu}$ are the gauge transformation
\begin{equation}
  \Delta = \frac{s_{12}(s_{13}+s_{23})}{4} \big(\Psi^{123} \big)^{2} +\frac{s_{12}s_{13}}{2} \Psi^{123} \Psi^{132} +  \frac{s_{13}(s_{12}+s_{23})}{4} \big(\Psi^{132}\big)^{2}\,,
\end{equation}
and
\begin{equation}
\begin{aligned}
  \Delta_{\mu} &=
   \frac{\left(s_{13}Q_{231}+s_{12}Q_{321}\right) \left(2P_{123}+2P_{132} + \epsilon_{2}{\cdot}\epsilon_{3}\right) \epsilon^1_{\mu }}{2 s_{23}}
  -\frac{\left( s_{13}Q_{231}+ s_{12}Q_{321}\right)^2 k^1_{\mu}}{2 s_{12} s_{13} s_{23}}
  \\
  &\quad + \text{cyclic}(1,2,3)\,.
\end{aligned}\label{}
\end{equation}

We may generalise the double copy in the rank-3 graviton current to higher rank. Recall that the graviton and gluon scattering amplitudes are associated with their BG currents as
\begin{equation}
\begin{aligned}
    M(1,2,\cdots,n+1) &= \lim_{s_{12\cdots n}\to0} s_{12\cdots n} \epsilon^{n+1}_{\mu\nu} \mathcal{J}^{123\cdots n}_{\mu\nu}\,, 
    \\ 
    A_{n}(1,2,\cdots,n+1) &= \lim_{s_{12\cdots n}\to0} s_{12\cdots n} \epsilon^{n+1}_{\mu} J^{12\cdots n}_{\mu} \,.
\end{aligned}\label{}
\end{equation}
This means that the gluon and graviton currents always have simple poles of $s_{12\cdots n}$, and the simple poles only contribute to the scattering amplitudes. We may expect the double copy structure even in the currents level. The $(n-2)!$-form of the KLT relation \cite{Bjerrum-Bohr:2010diw,Bjerrum-Bohr:2010mtb} is given by
\begin{equation}
  M_{n} =\lim_{k_{n}^{2} \rightarrow 0} \sum_{\rho, \tau \in S_{n-2}} \mathcal{A}_{n}^{\text {YM}}(1, \rho, n) \frac{S[\rho | \tau]_{1}}{s_{12} \cdots n-1} \mathcal{A}_{n}^{\text {YM}}(1, \tau, n)\,.
\label{KLT}\end{equation}
%

Since the KLT relation \eqref{KLT} is only guaranteed for the scattering amplitude level, the regular terms do not have to satisfy the KLT relation. Thus we propose the general KLT relation for the graviton currents as follows:
\begin{tcolorbox}
\begin{equation}
\begin{aligned}
  J_{\mu\nu}^{12\cdots n} = \sum_{\sigma,\tau\in S_{n-1}}J^{1\sigma}_{\mu} S[\sigma|\tau]_{1} J^{1\tau}_{\nu} + \text{regular terms}   + k_{\mu}^{\mathcal{P}} \Delta_{\nu} + \Delta_{\mu} k_{\nu}^{\mathcal{P}} + k^{\mathcal{P}}_{\mu} k^{\mathcal{P}}_{\nu} \Delta \,.
\end{aligned}\label{current_KLT}
\end{equation}
\end{tcolorbox}

\subsection{Currents in the spinor-helicity basis}
The off-shell currents are more complicated as the rank increases in general. In order to verify the current KLT relation \eqref{current_KLT} explicitly, we have to multiply two gluon currents with the $\big((n-1)!\big)^{2}$ different combinations and add all of them by multiplying the elements of the KLT kernel $S[\alpha|\beta]_{1}$ in a specific way. The checking process requires huge calculations indeed. Thus it is essential to reduce the number of terms in each gluon current, so that we employ the spinor-helicity formalism for the rank-4 currents. In this formalism, a proper choice of reference momenta of polarization vectors/tensors greatly simplifies the form of the currents. There are two distinct classes in helicity choice for the graviton and gluon currents:
\begin{enumerate}
	\item MHV currents: $(1+,2+,3+,\cdots)$ and $(1-,2+,3+,\cdots)$ 
	\item More than two opposite helicities : $(1-,2-,3+,\cdots)$ and $(1-,2-,3-,4+,\cdots)$ etc
\end{enumerate} 
Usually the term `MHV' is typically used in scattering amplitudes with $n-2$ gluons/gravitons of positive helicity and two gluons/gravitons of negative helicity. Here we use the term `MHV currents' for all positive helicity or with $n-1$ positive helicity and one negative helicity, which is related with the MHV amplitudes. In the MHV current case, we can choose reference momenta so that all the polarization vectors/tensors are orthogonal to each other $\epsilon^{i}\cdot \epsilon^{j} = 0$. As we will show later, in this case all the currents become null and orthogonal to each other. Then the graviton and gluon recursions reduce to a simpler form as in \eqref{reduced_recursions}. On the other hand, we have to consider the full recursion relation for the second class. 

 The helicity-2 polarization tensors $\varepsilon_{\mu\nu}^{i}$ for gravitons are given by a simple product of two polarization vectors, 
\begin{equation}
  \varepsilon^{i+}_{\mu\nu} = \epsilon^{i+}_{\mu} \epsilon^{i+}_{\nu}\,, \qquad \varepsilon^{i-}_{\mu\nu} = \epsilon^{i-}_{\mu} \epsilon^{i-}_{\nu}\,.
\end{equation}
The polarization vectors for the on-shell states satisfy
\begin{equation}
  \epsilon^{i} \cdot \epsilon^{i} = 0\,, \qquad k^{i} \cdot \epsilon^{i} = 0\,.
\label{on-shell_polarization}\end{equation}
Note that any four-dimensional vector can be represented by a pair of Weyl spinors using the gamma matrices, $p_{\mu} \gamma^{\mu}=-|p\rangle[p|+| p]\langle p|$. Then the polarization vectors with external momenta $k^{i}$, $\epsilon^{i}_{\mu}(k^{i})$,  are written by
\begin{equation}
	\epsilon^{i-}_{\mu}(k^{i} ; q)=-\frac{\left\langle i\left|\gamma^{\mu}\right| q\right]}{\sqrt{2}[q i]}, 
	\qquad 
	\epsilon^{i+}_{\mu}(k^{i} ; q)=-\frac{\left\langle q\left|\gamma^{\mu}\right| i\right]}{\sqrt{2}\langle q i\rangle}\,,
\label{}\end{equation}
where $q_{\mu}$ is an auxiliary null vector, called the reference momentum, reflecting the gauge symmetry of the massless vector fields. This representation satisfies the on-shell conditions \eqref{on-shell_polarization} automatically. 
We further introduce the generalised Mandelstam variables
\begin{equation}
  s_{i j}=-\left(k_{i}+k_{j}\right)^{2}, \quad s_{i j k}=-\left(k_{i}+k_{j}+k_{k}\right)^{2}, \quad \text { etc }
\label{}\end{equation}
These are represented by spinors by the Fierz identity, $\left\langle 1\left|\gamma^{\mu}\right| 2\right]\left\langle 3\left|\gamma_{\mu}\right| 4\right]=2\langle 13\rangle[24]$,
\begin{equation}
  s_{ij}= -\left\langle ij\right\rangle[ij] \,.
\label{}\end{equation}
%

\subsection{Class 1: MHV currents}

Here we consider the MHV currents: all positive helicity or one negative helicity. Without loss of generality we make a choice of the reference momenta as follows:
\begin{equation}
\begin{aligned}
  (1+,2+,3+,\cdots) &: \quad  \text{for all } \epsilon^{i+}_{\mu}, \text{we choose the same }  q_{\mu}
  \\
  (1-,2+,3+,\cdots) &: \quad \begin{cases} \text{for}~ \epsilon^{1-}_{\mu}\,,~ q_{\mu} = k^{2}_{\mu}
  \\
  \text{for}~ \epsilon^{i+} \,,~ q_{\mu} = k^{1}_{\mu}\,,\quad i>1
   \end{cases}  \,.
\end{aligned}\label{}
\end{equation}
Under this choice, the polarization vectors are orthogonal to each other. Since the graviton polarization tensors consist with the polarization vector $\mathcal{E}^{i}_{\mu\nu} = \epsilon^{i}_{\mu} \epsilon^{i}_{\nu}$, the polarization tensors are also orthogonal to each other
\begin{equation}
  \varepsilon^{i}_{\mu\nu} \varepsilon^{j}_{\nu\rho} = 0 \,, \quad \text{for all } i, j\,.
\label{orthogonal_graviton_polarization}\end{equation}

In this case the graviton recursion admits a special subsector. Using the nilpotency and orthogonality of the polarization tensors, one can show that all the MHV graviton currents are orthogonal to each other, $J^{\mathcal{P}}_{\mu\nu} J^{\mathcal{Q}}_{\nu \rho} = 0$, where $\mathcal{Q}$ and $\mathcal{R}$ are arbitrary words. From this property, the recursion relation for the MHV currents reduces to 
\begin{equation}
\begin{aligned}
  \mathcal{J}^{\mathcal{P}}_{\mu\nu} &= 
  - \frac{1}{s_{\mathcal{P}}} \sum_{ \mathcal{P} = \mathcal{Q} \cup \mathcal{R} } \bigg( \big(k^{\mathcal{R}}_{\rho} \mathcal{J}^{\mathcal{Q}}_{\rho\sigma} k^{\mathcal{R}}_{\sigma}\big)\mathcal{J}^{\mathcal{R}}_{\mu\nu} 
  - \big(\mathcal{J}^{\mathcal{Q}}_{\mu\rho} k^{\mathcal{R}}_{\rho} \big) \big( \mathcal{J}^{\mathcal{R}}_{\nu\sigma}k^{\mathcal{Q}}_{\sigma}\big) \bigg)\,.
\end{aligned}\label{reduced_recursions}
\end{equation}
We can prove the orthogonality between the MHV currents by induction. By assumption, the rank-1 currents identified with the polarization tensors are orthogonal to each other \eqref{orthogonal_graviton_polarization}. Next, let us assume that all the currents up to rank-$p$ are orthogonal to each other
\begin{equation}
  \mathcal{J}^{\mathcal{Q}}_{\mu\nu} \mathcal{J}^{\mathcal{R}}_{\nu\rho} =0\,, \qquad |\mathcal{Q}|\leq p ~ \text{and} ~ |\mathcal{R}|\leq p\,.
\label{orthogonality_of_MHV}\end{equation}
Under this condition, the graviton recursion relation reduces to \eqref{reduced_recursions} because all the terms including contractions between the graviton currents vanish. Then, from \eqref{reduced_recursions} and \eqref{orthogonality_of_MHV}, we can show that all the rank-$(p+1)$ currents, $\mathcal{J}^{\mathcal{P}}_{\mu\nu}$ and $\mathcal{J}^{\mathcal{P}'}_{\nu\rho}$ with $|\mathcal{P}|=|\mathcal{P}'|= (p+1)$, are orthogonal with the lower rank currents $\mathcal{J}^{\mathcal{P}}_{\mu\nu} \mathcal{J}^{\mathcal{Q}}_{\nu\rho} = 0$, where $|\mathcal{Q}| \leq p$ and themselves, $\mathcal{J}^{\mathcal{P}}_{\mu\nu} \mathcal{J}^{\mathcal{P}'}_{\nu\rho} = 0$. 

A corollary of the MHV recursion relation \eqref{reduced_recursions} is that the MHV graviton currents are proportional to the polarization vectors. In other words the graviton currents do not contain terms that are proportional to the external momenta $k^{i}_{\mu}$. One can show this by induction as before. The rank-1 current is trivial. If we assume that the up to rank-n currents are written in the following form
\begin{equation}
  \mathcal{J}^{\mathcal{P}}_{\mu\nu} = \sum_{i\in\mathcal{P}} \mathcal{J}^{\mathcal{P}}_{ij} \epsilon^{i}_{(\mu}\epsilon^{j}_{\nu)}\,, \qquad |\mathcal{P}|\leq n \,,
\label{MHV_current}\end{equation}
and substitute into the recursion relation for the rank-$(n+1)$ current \eqref{reduced_recursions},  we can easily show that the rank-$(n+1)$ current can be also written in the same form as \eqref{MHV_current}.

So far we have discussed only for the graviton currents, however the MHV gluon currents also satisfy the same properties \cite{Berends:1987me},
\begin{equation}
  J^{P}_{\mu} J^{Q}_{\mu} = 0\,, \qquad J_{\mu}^{P} = \sum_{i} J^{P}_{i} \epsilon^{i}_{\mu}\,,
\label{}\end{equation}
and the recursion relation is reduced into a simple form 
\begin{equation}
  J^{P}_{\mu} = \frac{i}{s_{P}} \sum_{P=QR} \Big( \big( J^{Q}\cdot k^{R}\big) J_{\mu}^{R} -\big(J^{R}\cdot k^{Q}\big) J^{Q}_{\mu}\Big)\,.
\label{MHV_recursion_gluon}\end{equation}
Note that the MHV currents for gluon and graviton both cannot have the gauge transformation terms because the MHV recursions \eqref{reduced_recursions} and \eqref{MHV_recursion_gluon} do not have terms proportional to the $k^{P}_{\mu}$ or $k^{\mathcal{P}}_{\mu}$ respectively. Further we propose the exact KLT relation for the MHV currents even without any discrepancy in regular terms
\begin{equation}
  \mathcal{J}^{\mathcal{P}}_{\mu\nu} = \sum_{\alpha,\beta\in S_{\mathcal{P}^{*}}} J^{1\alpha}_{\mu} S[\alpha|\beta]_{1} J^{1\beta}_{\nu}\,,
\label{exact_KLT}\end{equation}
where $S_{\mathcal{P}^{*}}$ is a permutation group for the word $\mathcal{P}^{*} = \{\mathcal{P}_{2},\mathcal{P}_{3},\cdots ,\mathcal{P}_{|P|}\}$.

Let us compute the MHV graviton currents and examine the exact current KLT relation up to rank-4. 
\\~\\
$\bullet$ \textbf{ Rank-2}
\\
If we compute the rank-2 MHV graviton currents from the recursion \eqref{reduced_recursions}, we have
\begin{equation}
\begin{aligned}
  J_{\mu\nu}^{12} = -\frac{1}{s_{12}} \Big( (k^{1}\cdot \epsilon^{2})^{2} \epsilon^{1}_{\mu}\epsilon^{1}_{\nu} 
  - (k^{2}\cdot \epsilon^{1}) (k^{1}\cdot \epsilon^{2}) \epsilon^{1}_{(\mu}\epsilon^{2}_{\nu)} + (1\leftrightarrow 2)\Big)\,.
\end{aligned}\label{}
\end{equation}
Similarly the rank-2 MHV gluon currents are
\begin{equation}
\begin{aligned}
  J^{12}_{\mu} = -\frac{\sqrt{2}}{s_{12}} \Big((k^{2}\cdot \epsilon^{1})\epsilon^{2}_{\mu} - (k^{1}\cdot \epsilon^{2})\epsilon^{1}_{\mu} \Big)\,.
\end{aligned}\label{}
\end{equation}
It is straightforward to show that these currents satisfy the current KLT relation
\begin{displaymath}
  \mathcal{J}^{12}_{\mu\nu} = J_{\mu}^{12} S[2|2]_{1} J_{\nu}^{12}\,,
\end{displaymath}
where $S[2|2]_{1} = -\frac{1}{2} s_{12}$.
\\~\\
$\bullet$ \textbf{Rank-3}
\\
The rank-3 graviton BG current is obtained from the recursion relation \eqref{reduced_recursions}
\begin{equation}
    \mathcal{J}_{\mu\nu}^{123} = - \frac{1}{2} \frac{1}{s_{123}} k^{123}_{\rho} k^{123}_{\sigma} \bigg[\ 
    \mathcal{J}^{12}_{\rho\sigma} \mathcal{J}^{3}_{\mu\nu} - \mathcal{J}^{12}_{\rho\mu} \mathcal{J}^{3}_{\sigma\nu}
  + \mathcal{J}^{3}_{\rho\sigma} \mathcal{J}^{12}_{\mu\nu} - \mathcal{J}^{3}_{\rho\mu} \mathcal{J}^{12}_{\sigma\nu} + \text{Perm [1,2,3]}\ \bigg]\,.
\label{}\end{equation}
Using the KLT relation for the rank-2 currents, we have
\begin{equation}
\begin{aligned}
  \mathcal{J}^{123}_{\mu\nu}  &= \frac{s_{12}}{2s_{123}} \bigg[
  \Big( (k^{3}\cdot J^{12})\epsilon^{3}_{\mu} {-} (k^{12}\cdot\epsilon^{3})J^{12}_{\mu}\Big)\Big( (k^{3}\cdot J^{12})\epsilon^{3}_{\nu} - (k^{12}\cdot\epsilon^{3})J^{12}_{\nu}\Big)
  \ \bigg] {+} \text{cyclic[1,2,3]} \,.
\end{aligned}\label{rank3_KLT}
\end{equation}

For simplicity we introduce a new current, $\hat{J}^{ij,k}_{\mu} = (k^{k}\cdot J^{ij})\epsilon^{k}_{\mu} - (k^{ij}\cdot \epsilon^{k})J^{ij}_{\mu}$. This current is antisymmetric under the interchange of $i$ and $j$, $\hat{J}^{ij,k}_{\mu} = -\hat{J}^{ji,k}_{\mu}$, due to the property of the rank-2 currents that $J^{ij}_{\mu} =- J^{ji}_{\mu}$. Then the rank-3 gluon currents are recast
\begin{equation}
\begin{aligned}
  J^{ijk}_{\mu} = -\frac{\sqrt{2}}{s_{ijk}} \bigg[ \hat{J}^{ij,k}_{\mu} - \hat{J}^{jk,i}_{\mu} \bigg]\,, 
  \qquad 
  J^{ikj}_{\mu} = -\frac{\sqrt{2}}{s_{ijk}} \bigg[ \hat{J}^{ik,j}_{\mu} + \hat{J}^{jk,i}_{\mu} \bigg]\,.
\end{aligned}\label{}
\end{equation}
Conversely, $\hat{J}^{ij,k}_{\mu}$ can be written in terms of the rank-3 currents
\begin{equation}
  \hat{J}^{ij,k}_{\mu} = \frac{1}{\sqrt{2}} \big(s_{ik} J_{\mu}^{kij} -s_{jk} J^{kji}_{\mu}\big)\,.
\label{hJ and J}\end{equation}
We then rewrite $\mathcal{J}^{123}_{\mu\nu}$ in \eqref{rank3_KLT} using the new currents
\begin{equation}
  \mathcal{J}^{123}_{\mu\nu} =\frac{1}{2 s_{123}} \Big( s_{12}\hat{J}^{12,3}_{\mu}\hat{J}^{12,3}_{\nu} +s_{23}\hat{J}^{23,1}_{\mu}\hat{J}^{23,1}_{\nu} +s_{13}\hat{J}^{13,2}_{\mu}\hat{J}^{13,2}_{\nu}\Big)\,.
\label{hJ3}\end{equation}
Comparing this with the exact KLT relation \eqref{exact_KLT}, we have
\begin{equation}
\begin{aligned}
  \mathcal{J}^{123}_{\mu\nu}  &= J^{123}_{\mu} S[23|23]_{1} J^{123}_{\nu} +2J^{123}_{(\mu} S[23|32]_{1} J^{132}_{\nu)} + 
  J^{132}_{\mu} S[32|32]_{1} J^{132}_{\nu}
  \\
  &\quad -\frac{1}{2s^{2}_{123}}\Big(s^{12}\hat{J}^{12,3}_{\mu} - s^{13}\hat{J}^{13,2}_{\mu}+ s^{23}\hat{J}^{23,1}_{\mu}\Big)\Big(s^{12}\hat{J}^{12,3}_{\nu} -s^{13}\hat{J}^{13,2}_{\nu}+s^{23}\hat{J}^{23,1}_{\nu}\Big)\,.
\end{aligned}\label{}
\end{equation}
Using the identity \eqref{hJ and J}, we can rewrite the parenthesis in the last line reduces to
\begin{equation}
\begin{aligned}
  \frac{1}{2\sqrt{2} s_{123}} \Big(
    s_{12} s_{13} \big(J^{312}_{\mu}-J^{213}_{\mu}\big) \Big)
  + s_{12} s_{23} \big(J^{123}_{\mu}-J^{321}_{\mu}\big) \Big)
  + s_{13} s_{23} \big(J^{231}_{\mu}-J^{132}_{\mu}\big) \Big) = 0 \,,
\end{aligned}\label{}
\end{equation}
and it vanishes due to the reflection identity for the rank-3 gluon currents
\begin{equation}
  J_{\mu}^{ijk} = J_{\mu}^{kji}\,.
\label{}\end{equation}
This shows the rank-3 graviton current $\mathcal{J}^{ijk}_{\mu\nu}$ satisfies the exact KLT relation.
~\\~\\
$\bullet$ \textbf{Rank-4}\\
The structure of the rank-4 currents are similar to the rank-3 currents. From the recursion, the rank-4 graviton currents are given by
\begin{equation}
    \mathcal{J}^{1234}_{\mu\nu} =-\frac{2}{s_{1234}} k^{1234}_{\rho} k^{1234}_{\sigma} \bigg[\ 
    \frac{1}{3!} J^{123}_{\rho[\sigma} J^{4}_{\mu]\nu} 
  + \frac{1}{4}  J^{12}_{\rho[\sigma} J^{34}_{\mu]\nu} 
  + \frac{1}{3!} J^{4}_{\rho[\sigma} J^{123}_{\mu]\nu} \ \bigg]
  + \text{Perm}[1,2,3,4] \,,
\label{J4}\end{equation}
We introduce a pair of currents generalising the rank-3 counterpart $\hat{J}_{\mu}^{ij,k}$ \eqref{hJ3}
\begin{equation}
\begin{aligned}
  \hat{J}^{ijk,l}_{\mu} &=(k^{k}\cdot J^{ijk} ) J^{l}_{\mu} - (k^{ijk}\cdot J^{l} )J^{ijk}_{\mu}\,,
  \\
  \hat{J}^{ij,kl}_{\mu} &=(k^{kl}\cdot J^{ij} ) J^{kl}_{\mu} - (k^{ij}\cdot J^{kl} )J^{ij}_{\mu}\,.
\end{aligned}\label{hJ4}
\end{equation}
Then the graviton current \eqref{J4} can be recast in the following form
\begin{equation}
\begin{aligned}
  \mathcal{J}^{1234}_{\mu\nu} = -\frac{1}{s_{1234}} \bigg[& 
    \sum_{\alpha,\beta\in S_{\{2,3\}}} S[\alpha|\beta]_{1} \hat{J}^{1\alpha,4}_{\mu} \hat{J}^{1\beta,4}_{\nu} 
  + \sum_{\alpha,\beta\in S_{\{2,4\}}} S[\alpha|\beta]_{1} \hat{J}^{1\alpha,3}_{\mu} \hat{J}^{1\beta,3}_{\nu}
    \\
  &
  + \sum_{\alpha,\beta\in S_{\{3,4\}}} S[\alpha|\beta]_{1} \hat{J}^{1\alpha,2}_{\mu} \hat{J}^{1\beta,2}_{\nu}
 + \sum_{\alpha,\beta\in S_{\{3,4\}}} S[\alpha|\beta]_{2} \hat{J}^{2\alpha,1}_{\mu} \hat{J}^{2\beta,1}_{\nu}
   \\
  &
  +\frac{1}{4}s_{12}s_{34} \hat{J}^{12,34}_{\mu}\hat{J}^{12,34}_{\nu}
  +\frac{1}{4}s_{13}s_{24} \hat{J}^{13,24}_{\mu}\hat{J}^{13,24}_{\nu}
  +\frac{1}{4}s_{14}s_{23} \hat{J}^{14,23}_{\mu}\hat{J}^{14,23}_{\nu} \ \bigg]\,.
\end{aligned}\label{graviton_rank4}
\end{equation}
Similarly, the rank-4 gluon recursion is given by
\begin{equation}
\begin{aligned}
  J^{1234}_{\mu} = -\frac{\sqrt{2} }{s_{1234}} \bigg[\ & \big(J^{1}\cdot k^{234}\big) J^{234}_{\mu} - \big(J^{234}\cdot k^{1}\big) J^{1}_{\mu} 
 +\big(J^{12}\cdot k^{34}\big) J^{34}_{\mu} - \big(J^{34}\cdot k^{12}\big) J^{12}_{\mu}
  \\
  & +\big(J^{123}\cdot k^{4}\big) J^{4}_{\mu} - \big(J^{4}\cdot k^{123}\big) J^{123}_{\mu}
  \ \bigg] \,,
\end{aligned}\label{}
\end{equation}
and we can recast the current using \eqref{hJ4}
\begin{equation}
\begin{aligned}
  J^{1234}_{\mu} = - \frac{\sqrt{2}}{s_{1234}} \bigg[ \hat{J}^{123,4}_{\mu} - \hat{J}^{234,1}_{\mu}
  + \hat{J}^{12,34}_{\mu}\bigg]\,,
\end{aligned}\label{JinhJ}
\end{equation}
or conversely
\begin{equation}
\begin{aligned}
  \hat{J}^{123,4}_{\mu} &= \frac{1}{\sqrt{2}}\left(s_{14} J_{\mu}^{4123}-s_{24} J_{\mu}^{4213}-s_{24} J_{\mu}^{4231}-s_{34} J_{\mu}^{1234}\right)\,,
  \\
  \hat{J}^{12,34}_{\mu} &= \frac{1}{\sqrt{2}}\left(s_{13} J_{\mu}^{2134}-s_{14} J_{\mu}^{2143}+s_{24} J_{\mu}^{1243}-s_{23} J_{\mu}^{1234}\right)\,.
\end{aligned}\label{}
\end{equation}
Substituting \eqref{JinhJ} into the exact current KLT relation and comparing with the graviton current \eqref{graviton_rank4}, we have
\begin{equation}
\begin{aligned}
  \mathcal{J}^{1234}_{\mu\nu} &- \sum_{\alpha,\beta\in S_{\{2,3,4\}}} J^{1\alpha}_{\mu} S[\alpha|\beta]_{1} J^{1\beta}_{\nu} 
  \\
  &= \frac{1}{4 s_{1234}} \bigg[ 
   s_{12} s_{23} \hat{J}_{\mu}^{123,4}\hat{J}_{\nu}^{123,4} + \text{permutation}[1,2,3,4] 
  \\
  &\qquad\qquad\qquad + s_{12} s_{34} \hat{J}_{\mu}^{12,34} \hat{J}_{\nu}^{12,34} 
  + s_{13} s_{24} \hat{J}_{\mu}^{13,24} \hat{J}_{\nu}^{13,24} 
  + s_{14} s_{23} \hat{J}_{\mu}^{14,23} \hat{J}_{\nu}^{14,23} \bigg]\,,
  \\
  &\quad - \frac{1}{8} \bigg[\  s_{12} s_{23} s_{34} J^{1234}_{\mu} J^{1234}_{\nu} + \text{permutation}[1,2,3,4] 
  \\
  &\qquad\qquad + s_{12}\big(s_{13}s_{14} + s_{23}s_{24}\big) \big(J^{2134}_{\mu}+J^{2143}_{\mu}\big)\big(J^{2134}_{\nu}+J^{2143}_{\nu}\big)
  \\
  &\qquad\qquad +\Big( s_{13}s_{23}s_{34}\big(J^{1324}_{\mu}+J^{1342}_{\mu}\big)\big(J^{1324}_{\nu} +J^{1342}_{\nu}\big) + \big(3\leftrightarrow 4\big)\Big)\ \bigg]\,.
\end{aligned}\label{}
\end{equation}
One can show that the right hand side vanishes exactly, and this shows the exact current KLT relation for the rank-4 currents.

%
%
%
%

\subsection{Class 2: Two opposite helicities}
Let us consider the second class involving two opposite helicities. The first nontrivial currents in this class starts from rank-4, such as $(1-,2-,3+,4+)$. Our reference momenta convention is
\begin{equation}
\begin{aligned}
  \epsilon^{1}_{\mu\nu} : q^{1} = k^{3}\,,  \qquad  \epsilon^{2}_{\mu\nu} : q^{2}= k^{3}\,,
  \qquad
  \epsilon^{i}_{\mu\nu} : q^{i} = k^{1}\,, \quad \text{for}~i\geq 3\,.
\end{aligned}\label{}
\end{equation}
Using the convention, we have
\begin{equation}
\begin{aligned}
  k^{i} \cdot\epsilon^{1-} &= - \frac{s_{1i}[i3]}{\sqrt{2}[1i][31]} \,, 
  &\qquad 
  k^{i} \cdot\epsilon^{2-} &= - \frac{s_{i2}[i3]}{\sqrt{2}[2i][32]} \,, 
  \qquad 
  k^{i} \cdot\epsilon^{j+} = - \frac{s_{ij}\left\langle i1\right\rangle}{\sqrt{2}\left\langle ji\right\rangle\left\langle 1j\right\rangle} \,,
  \\
  \epsilon^{2-}\cdot \epsilon^{4+} &= \frac{\left\langle 12\right\rangle [34]}{\left\langle 14\right\rangle[23]}\,,
  &\qquad
  \epsilon^{2-}\cdot \epsilon^{i} &= 0\,,
\end{aligned}\label{}
\end{equation}
where $1 \leq i,j \leq 4$. Further we have the following identities:
\begin{equation}
\begin{aligned}
  (k^{1} \cdot \epsilon^{2}) (k^{3} \cdot \epsilon^{4} )  &= - \frac{s_{13}}{2} (\epsilon^{2} \cdot \epsilon^{4} ) \,,
  \\
  (k^{2} \cdot \epsilon^{4}) (k^{4} \cdot \epsilon^{2} )  &= - \frac{s_{24}}{2} (\epsilon^{2} \cdot \epsilon^{4} ) \,,
  \\
  ( k^{2} \cdot \epsilon^{3} ) ( k^{4} \cdot \epsilon^{1} ) &= ( k^{2} \cdot \epsilon^{1} ) ( k^{4} \cdot \epsilon^{3} )\,.
\end{aligned}\label{}
\end{equation}

We now present the rank-4 graviton current $\mathcal{J}_{\mu\nu}^{1234}$. It is straightforward to compute the explicit expression of the graviton current by solving the recursion, however, we just write down it using the current KLT relation because of the lengthy expression 
\begin{equation}
\begin{aligned}
  \mathcal{J}^{1234}_{\mu\nu}= \sum_{\alpha,\beta\in S_{\{2,3,4\}}} J_{\mu}^{1\alpha} S[\alpha|\beta]_{1} J_{\nu}^{1\beta} + 2k^{1234}_{(\mu} \Delta^{1234}_{\nu)} + k^{1234}_{\mu}k^{1234}_{\nu} \Delta^{1234} +\text{regular terms}\,.
\end{aligned}\label{}
\end{equation}
We may decompose the gauge transformations, $\Delta^{1234}_{\mu}$ and $\Delta^{1234}$, into two parts according to their origin,  from the gluon currents, $k^{1234}_{\mu} \Psi^{P}$, or the graviton current. (See appendix \ref{appendix_B} for the explicit form of $\Psi^{P}$) If we denote the contribution from the graviton currents as $\Phi^{1234}_{\mu}$ and $\Phi^{1234}$, the total gauge transformation are written as
\begin{equation}
\begin{aligned}
  \Delta_{\mu}^{1234} &= \Phi^{1234}_{\mu} + \sum_{\alpha,\beta\in S_{\{2,3,4\}}} \Psi^{1\alpha} S[\alpha|\beta]_{1} J_{\mu}^{1\beta}
  \\
  \Delta^{1234} &= \Phi^{1234} + \sum_{\alpha,\beta \in S_{\{2,3,4\}}} \Psi^{1\alpha} S[\alpha|\beta]_{1}\Psi^{1\beta}\,.
\end{aligned}\label{gauge_transform_rank4}
\end{equation}
where
\begin{equation}
\begin{aligned}
  \Phi^{1234} &= \frac{Q_{234,1}^2}{ 2 } 
 \Bigg[\frac{2(s_{24}-s_{13})}{s_{24}s_{124}s_{234}} 
  +s_{123} s_{134} \bigg( \frac{2 (s_{13} {-} s_{24})}{s_{24}} - \frac{(s_{14}{+}s_{24}) s_{13} (s_{23} + s_{24})}{s_{14} s_{23} s_{24}} + \frac{s_{12}s_{34}}{s_{14} s_{23}} \bigg)  
  \\ 
  &\qquad~  +\frac{s_{13} (s_{14} + s_{24}) (s_{23} +s_{24}) - s_{12} s_{24} s_{34}}{s_{124} s_{14} s_{23} s_{234} s_{24}}  
  {+} \frac{2}{s_{24}} \bigg(\frac{s_{24} {+} s_{34}}{s_{34}}\Big(\frac{s_{13}{+}s_{14}}{s_{12} s_{234}}\Big) {+}\frac{ s_{14} }{s_{12} s_{124}}\bigg) \Bigg]\,,
\end{aligned}\label{}
\end{equation}
and $\Psi^{P}$ and $\Phi^{1234}_{\mu}$ are in appendix \ref{appendix_C}. Here  $Q_{ijk,l} = Q_{ijk} (k_{k}{\cdot}\epsilon_{l})$. 

Finally, we comment on the ambiguity of the graviton off-shell currents. Since the on-shell amplitudes are completely independent of regular terms and gauge transformations, these can be added or subtracted to the off-shell currents, and the form of the currents is not unique. In section \ref{Sec:2}, we discussed that redundancies reside in the DFT EoM. Thus we extracted the essential part of EoM by using the background projection operators $\Theta_{\mu}{}^{M}$ and $\bar{\Theta}_{\mu}{}^{M}$. Even though our EoM \eqref{DFT_perturbative_eom} is compact and useful to solve it, however, it may not be optimized for describing the current KLT relation. If we compute the KLT relation using the off-shell currents from the redefined EoM, then the regular and gauge transformation terms would be reduced or disappeared.

\section{Conservation of the off-shell currents}\label{Sec:5}
One of the characteristic features of the BG current is off-shell conservation. It represents gauge invariance as the Ward identity of the on-shell scattering amplitudes and gives a non-trivial test of the off-shell currents. For instance, the gluon currents in the Lorentz gauge condition satisfy $k^{\mathcal{P}}\cdot J^{\mathcal{P}} = 0$ without using equations of motion. In this section, we will consider the conservation of the off-shell currents in a nonlinear gauge condition. We will show that the conservation should be hold up to the gauge condition we have imposed. As an example, we will examine the conservation of the gluon currents under the Gervais-Neveu gauge condition. We then show that conservation of our graviton currents satisfies in the same way. 

\subsection{Conservation of gluon currents in Gervais-Neveu gauge condition}
Before discussing the graviton current case, let us consider the gluon current conservation in the Gervais-Neveu (GN) gauge condition as a toy model. It is a nonlinear gauge choice in gauge field $\mathbb{A}_{\mu}$
\begin{equation}
  \partial^{\mu} \mathbb{A}_{\mu} = \frac{i}{\sqrt{2}} \mathbb{A}^{\mu} \mathbb{A}_{\mu}\,.
\label{GN_gauge}\end{equation}
The perturbiner expansion of $\mathbb{A}_{\mu}$ is still the same as \eqref{gluon_perturbiner_expansion} because it is independent of the gauge choice. Substituting the perturbiner expansion into the GN gauge condition \eqref{GN_gauge} gives
\begin{equation}
  k^{P}_{\mu} \tilde{J}^{P}_{\mu} = \frac{1}{\sqrt{2}} \sum_{P=QR} \tilde{J}^{Q} \cdot \tilde{J}^{R} \,,
\label{conservation_GN}\end{equation}
where $\tilde{J}^{P}_{\mu}$ represents the gluon currents in the GN gauge condition. This implies that the conservation of currents in a nonlinear gauge condition should be modified. In the Lorentz gauge fixing, the right hand side is trivial, $k^{P}_{\mu} J^{P}_{\mu} = 0$, and it is consistent with the usual conservation of currents. However, in GN gauge condition, there is no reason the right hand side of \eqref{conservation_GN} vanish in general. Then the current conservation holds up to the GN gauge condition. We will check \eqref{conservation_GN} explicitly by computing the $k^{P}_{\mu} \tilde{J}^{P}_{\mu}$ by computing the gluon currents by solving the recursion relation.

Under the GN gauge condition, the action is reduces to
\begin{equation}
\begin{aligned}
  \mathcal{L}_{\rm YM} = \Tr \big( - \frac{1}{2} \partial^{\mu} \mathbb{A}^{\nu} \partial_{\mu} \mathbb{A}_{\nu} - i \sqrt{2} \partial^{\mu} \mathbb{A}^{\nu} \mathbb{A}_{\nu} \mathbb{A}_{\mu} + \frac{1}{4} \mathbb{A}_{\mu} \mathbb{A}_{\nu} \mathbb{A}^{\mu} \mathbb{A}^{\nu} \big) \,,
\end{aligned}\label{}
\end{equation}
and the corresponding equations of motion is
\begin{equation}
\begin{aligned}
  &\Box \mathbb{A}_{\mu} + i \sqrt{2} \Big( [ \partial^{\nu} \mathbb{A}_{\mu} , \mathbb{A}_{\nu} ] -  \big( \partial_{\mu} \mathbb{A}^{\nu} \big) \mathbb{A}_{\nu} \Big) + [ \mathbb{A}_{\nu} , \mathbb{A}_{\mu} ] \mathbb{A}^{\nu} = 0 \,,
\end{aligned}\label{}
\end{equation}
or in terms of the field strength $\mathbb{F}_{\mu\nu} = \partial_{\mu} \mathbb{A}_{\nu} - \partial_{\nu} \mathbb{A}_{\mu} -\frac{i}{\sqrt{2}}[\mathbb{A}_{\mu},\mathbb{A}_{\nu}]$,
\begin{equation}
  \Box \mathbb{A}_{\mu} = i\sqrt{2} \Big(\mathbb{F}_{\mu\nu}\mathbb{A}^{\nu} + \mathbb{A}^{\nu} \partial_{\nu} \mathbb{A}_{\mu} \Big)\,.
\label{}\end{equation}
As before, if we substitute the perturbiner expansion into the above EoM, we get the recursion relation for the off-shell gluon currents
\begin{equation}
  \tilde{J}^{P}_{\mu} = \frac{i\sqrt{2}}{s_{P}} \sum_{P=QR} \Big(\tilde{F}_{\mu\nu}^{Q} \tilde{J}_{\nu}^{R} + i\big(\tilde{J}^{Q} \cdot k^{R}\big) \tilde{J}^{R}_{\mu}\Big)\,,
\label{recursion_GN}\end{equation}
where $\tilde{J}$ and $\tilde{F}$ represent the currents in the GN gauge choice. The recursion relation of $\tilde{F}^{P}_{\mu\nu}$ is the same as the $F^{P}_{\mu\nu}$ in \eqref{YM_recursions}.

Let us now compute the currents $\tilde{J}_{\mu}^{P}$ explicitly using \eqref{recursion_GN} and examine the current conservation. The rank-2 current is given by
\begin{equation}
  \tilde{J}^{ij}_{\mu} = \frac{\sqrt{2}}{s_{ij}} \bigg[ ( k_{i} {\cdot} \epsilon_{j} ) \epsilon^{i}_{\mu} - ( k_{j} {\cdot} \epsilon_{i} ) \epsilon^{j}_{\mu} - ( \epsilon_{i} {\cdot} \epsilon_{j} ) k^{i}_{\mu} \bigg]\,,
\end{equation}
and comparing the Lorentz gauge result, the difference is the gauge transformation
\begin{equation}
  J^{ij}_{\mu\nu} - \tilde{J}^{ij}_{\mu\nu} = \frac{\big(\epsilon_{i}{\cdot}\epsilon_{j}\big) k^{ij}_{\mu}}{\sqrt{2} s_{ij}} \,.
\label{}\end{equation}
The conservation of $\tilde{J}^{ij}_{\mu}$ is 
\begin{equation}
  k^{ij}_{\mu} \tilde{J}^{ij}_{\mu} = \frac{\epsilon_{i}{\cdot}\epsilon_{j}}{\sqrt{2}} = \frac{1}{\sqrt{2}} \tilde{J}^{i} \cdot \tilde{J}^{j}\,,
\label{}\end{equation}
and this exactly reproduces the GN gauge condition \eqref{conservation_GN}.

Next, the rank-3 gluon current is 
\begin{equation}
\begin{aligned}
  \tilde{J}_{\mu}^{ijk} &= \frac{1}{s_{ijk}}\bigg[\ \frac{ \left(s_{ij}+s_{jk}\right) \left(2 P_{ijk} +\epsilon_{j}{\cdot}\epsilon_{k} \right)-2 P_{ikj} s_{ik} }{s_{jk}} \epsilon^i_{\mu}
  - 2 \Big(P_{jik}+P_{jki}+\frac{\epsilon_{i}{\cdot}\epsilon_{k}}{2}\Big) \epsilon^j_{\mu}
  \\
  &\qquad\qquad -\frac{s_{ik} \left(2 P_{jik}+\epsilon_{j}{\cdot}\epsilon_{k}\right) -2 P_{kji} \left(s_{ij}+s_{jk}\right)}{s_{ij}} \epsilon^k_{\mu }
  \\
  &\qquad\qquad +\frac{2 \left(s_{ij}T_{kji} -s_{jk} T_{ijk}\right)k^i_{\mu } +2\big(s_{jk} T_{jik} +s_{ij} S^{+}_{jik}\big)k^j_{\mu }}{s_{12} s_{23} }\ \bigg]\,.
\end{aligned}\label{}
\end{equation}
Again, the difference with the rank-3 gluon currents in the Lorentz gauge is as follows
\begin{equation}
\begin{aligned}
  &J^{ijk}_{\mu}-\tilde{J}^{ijk}_{\mu} = -\frac{k^{ijk}_{\mu }}{s_{ijk}} \left(\frac{S^+_{jik}+2 T_{kji}}{2 s_{jk}} +\frac{2 T_{jik}-S^+_{jki}}{2 s_{ij}}\right)+\frac{(\epsilon_{i}{\cdot}\epsilon_{j}) \epsilon^{k}_{\mu }}{2 s_{ij}}-\frac{(\epsilon_{j}{\cdot}\epsilon_{k})\epsilon ^i_{\mu }}{2 s_{jk}}\,,
\end{aligned}\label{}
\end{equation}
where the first term on the right hand side is a gauge transformation and the second term is a regular term. 
If we contract $k^{ijk}_{\mu}$ with $\tilde{J}^{ijk}_{\mu}$, we get
\begin{equation}
  k^{ijk}_{\mu} \tilde{J}^{ijk}_{\mu} = -\frac{T_{jik}}{s_{ij}} - \frac{T_{kji}}{s_{jk}} = \frac{1}{\sqrt{2}} ( \tilde{J}^{ij}_{\mu} \tilde{J}^{k}_{\mu} + \tilde{J}^{i}_{\mu} \tilde{J}^{jk}_{\mu} ) \,.
\end{equation}
This is the GN gauge condition as we expected.

\subsection{Conservation of the graviton currents}
We now consider the graviton currents case. We have introduced the gauge choice in undoubled $D$-dimensional form  \eqref{gauge_condition_d}. As we have seen in the GN gauge choice, we get the conservation of graviton currents by substituting the perturbiner expansion \eqref{perturbiner_expansion_E} and \eqref{perturbiner_expansion_aux} into the gauge condition
\begin{equation}
  k^{\mathcal{P}} _{\nu}  \mathcal{J}^{\mathcal{P}}_{\nu \mu}  = k^{\mathcal{P}} _{\nu}  \Delta^{\mathcal{P}}_{\nu \mu}- 2  k^{\mathcal{P}}_{\mu} d^{\mathcal{P}}
  + 2 \sum_{\mathcal{P}=\mathcal{Q} \cup \mathcal{R}} (\mathcal{J}^{\mathcal{Q}}_{\mu \nu} -\varDelta^{\mathcal{Q}}_{\mu\nu} )k^{\mathcal{R}}_{\nu} d^{\mathcal{R}} \,,
\label{conserv_grav_current}\end{equation}
where $d^{\mathcal{P}}$ is the dilaton currents generated from the perturbiner expansion of $d$
\begin{equation}
  d = \sum_{\mathcal{P}} d^{\mathcal{P}} e^{ik_{\mathcal{P}}\cdot x}\,. 
\label{}\end{equation}
The EoM of the DFT dilaton $d$ is given by the generalised Ricci scalar $\mathcal{R}$ \eqref{DFT_eom_gauge_condition}. If we substitute the perturbation of the generalised metric \eqref{perturbed_H}, we have
\begin{equation}
\begin{aligned}
  \Box d &= -\frac{1}{4} \big(X^{\mu}{}_{\mu} + Y^{\mu}{}_{\mu}\big) 
  + \frac{1}{4}\mathcal{E}^{\mu \nu} \big( 4 \partial_{\mu} \partial_{\nu} d +X_{\mu\nu} -Y_{\mu\nu}\Big)
  - \frac{1}{4}\varDelta^{\mu \nu} \big( 4 \partial_{\mu} \partial_{\nu} d +X_{\mu\nu} +Y_{\mu\nu}\Big)\,,
\end{aligned}\label{eom_d}
\end{equation}
where $X_{\mu\nu}$ and $Y_{\mu\nu}$ are the auxiliary fields introduced in \eqref{DFT_perturbative_eom}. Then the recursion relation for $d^{\mathcal{P}}$ is derived from the EoM of $d$ \eqref{eom_d}
\begin{equation}
\begin{aligned}
  d^{\mathcal{P}} &= - \frac{1}{4s_{\mathcal{P}}}\bigg[\ \mathcal{X}^{\mathcal{P}}+\mathcal{Y}^{\mathcal{P}}
  -\sum_{\mathcal{P}=\mathcal{Q}\cup \mathcal{R}} \mathcal{J}^{\mathcal{Q}}_{\mu \nu} \Big( \mathcal{X}^{\mathcal{R}}_{\mu\nu} -\mathcal{Y}^{\mathcal{R}}_{\mu\nu} - 4 k^{\mathcal{R}}_{\mu} k^{\mathcal{R}}_{\nu} d^{\mathcal{R}}\Big)  
  \\
  &\qquad \qquad\quad
  +\sum_{\mathcal{P}=\mathcal{Q}\cup \mathcal{R}} \varDelta^{\mathcal{Q}}_{\mu\nu} \big(\mathcal{X}^{\mathcal{R}}_{\mu\nu}+\mathcal{Y}^{\mathcal{R}}_{\mu\nu} -4k^{\mathcal{R}}_{\mu} k^{\mathcal{R}}_{\nu} d^{\mathcal{R}}\big) \ \bigg] \,,
\end{aligned}
\end{equation}
where $\mathcal{X}^{\mathcal{P}}$ and $\mathcal{Y}^{\mathcal{P}}$ are the trace of $\mathcal{X}^{\mathcal{P}}_{\mu\nu}$ and $\mathcal{Y}^{\mathcal{P}}_{\mu\nu}$ respectively. Note that the rank-1 dilaton currents $d^{i}$ vanish because $\mathcal{X}_{\mu\nu}^{i} = \mathcal{Y}^{i}_{\mu\nu} =0$, and it starts from the rank-2.

We now compute the left hand side of \eqref{conserv_grav_current} explicitly using the graviton currents and verify the right hand side. Up to rank-3, we do not impose helicity on the polarization tensor. Let's consider the rank-1 conservation equation. It is ensured by the transverse condition of the polarization tensor
\begin{equation}
  k^{i} _{\nu}  \mathcal{J}^{i}_{\nu \mu} = k^{i} _{\nu} \varepsilon^{i}_{\nu \mu} = 0 \,.
\end{equation} 

Next, we compute the off-shell conservation for the rank-2 graviton currents $k^{ij}_{\nu}  \mathcal{J}^{ij}_{\nu \mu}$. If we use the explicit form of the rank-2 graviton current \eqref{rank-2_graviton_current}, the conservation equation reduces to
\begin{equation}
\begin{aligned}
  k^{ij} _{\nu}  \mathcal{J}^{ij}_{\nu \mu} & = 
  \frac{( \epsilon_{i} {\cdot} \epsilon_{j} )}{2}  \Big[ ( k_{i} {\cdot} \epsilon_{j} )  \epsilon^{i}_{\mu} + ( k_{j} {\cdot} \epsilon_{i} ) \epsilon^{j}_{\mu} \Big]
 - \bigg(\frac{( k_{i} {\cdot} \epsilon_{j} ) ( k_{j} {\cdot} \epsilon_{i} )  ( \epsilon_{i} {\cdot} \epsilon_{j} ) }{ s_{ij}} + \frac{( \epsilon_{i} {\cdot} \epsilon_{j} )^{2}}{4} \bigg) k^{ij}_{\mu} \,.
\label{rank2con}
\end{aligned}
\end{equation}
One can show that the first and the second terms on the right hand side are the same as $ k^{ij}_{\nu} \Delta^{ij}_{\nu \mu} $ and $ -2 k^{ij}_{\mu} d^{ij} $, respectively. Thus it satisfies the rank-2 gauge fixing condition
\begin{equation}
  k^{ij} _{\nu} \mathcal{J}^{ij}_{\nu \mu} = k^{ij} _{\nu}  \Delta^{ij}_{\nu \mu}- 2  k^{ij}_{\mu} d^{ij} \,.
\end{equation}
Finally, let's consider the rank-3 case. The rank-3 gauge fixing condition is written by
\begin{equation}
  k^{ijk} _{\nu}  \mathcal{J}^{ijk}_{\nu \mu}  = k^{ijk}_{\nu}  \Delta^{ijk}_{\nu \mu}- 2  k^{ijk}_{\mu} d^{ijk}+ \sum_{\text{Perm}[ijk]} \mathcal{J}^{i}_{\mu \nu}  k^{jk}_{\nu} d^{jk} \,.
\label{rank3_conserv}\end{equation}
We compute each term on the right hand side and compare with the left hand side. The first terms is
\begin{equation}
\begin{aligned}
  k^{ijk}_{\nu}  \Delta^{ijk}_{\nu \mu} &=  \Bigg[ 
  \frac{\epsilon_{j} {\cdot} \epsilon_{k}}{8} \Big(2 Q_{jki} - Q_{jik} 
  + \frac{s_{ij}\big( 2 Q_{kji} - Q_{kij}\big)}{s_{jk}} -2P_{jki} ( k_{k} {\cdot} \epsilon_{j} ) \Big)- \frac{P_{ijk} T_{ijk}}{2} 
  \\
  & \qquad
  - \frac{( k_{i} {\cdot} \epsilon_{j} )( k_{j} {\cdot} \epsilon_{k} ) T_{jki}  }{2s_{jk}} +(j\leftrightarrow k)\Bigg] \epsilon^{i}_{\mu}
+ \Bigg[ \frac{S^{+}_{ikj}\big( Q_{jki} - 2 Q_{jik} \big)}{4s_{ij}} +(j\leftrightarrow k) \Bigg] k^{i}_{\mu}
\\
& \quad
+ \text{cyclic}[i,j,k]\,,
\end{aligned}
\end{equation}
the second term is
\begin{equation}
\begin{aligned}
  -2 d^{ijk} k^{ijk}_{\mu} &=  \frac{1}{s_{ijk}} \Bigg[ \frac{( Q_{ijk} + Q_{jki} + Q_{kij} )^{2}}{12} 
  + \frac{s_{ik}}{s_{ij}} \Bigg(\frac{(T_{ijk})^{2}}{4} +P_{ikj} S^{+}_{ikj} (k_{j} {\cdot} \epsilon_{i})\Bigg)
\\
& \qquad\qquad
+ P_{ijk} R_{ijk} (k_{k} {\cdot} \epsilon_{i}) +(i\leftrightarrow j)\Bigg] k^{ijk}_{\mu} + \text{cyclic}[i,j,k]\,,
\end{aligned}
\end{equation}
and the last term is 
\begin{equation}
\begin{aligned}
  \sum_{\text{Perm}[ijk]} \mathcal{J}^{i}_{\mu \nu}  k^{jk}_{\nu} d^{jk} & =  \frac{\epsilon_{j} {\cdot} \epsilon_{k}}{4} \Big( S^{+}_{jik} + 4P_{jki} ( k_{k} {\cdot} \epsilon_{j} )\Big) \epsilon^{i}_{\mu} + \text{cyclic}[i,j,k]\,.
\end{aligned}
\end{equation}
If we compare these results with the left hand side of \eqref{rank3_conserv}, the gauge condition is exactly satisfied. This shows the rank-3 graviton off-shell conservation. We can continue this job to any higher rank currents. 

As a final comment, we consider the conservation of the MHV case, $k^{\mathcal{P}}_{\mu} \mathcal{J}_{\mu\nu}^{\mathcal{P}} = 0$. By substituting the graviton currents \eqref{reduced_recursions}, we can show that the MHV currents trivially satisfy the off-shell conservation exactly
\begin{equation}
\begin{aligned}
  k^{\mathcal{P}}_{\mu} \mathcal{J}_{\mu\nu}^{\mathcal{P}} &= - \frac{1}{s^{\mathcal{P}}} k^{\mathcal{P}}_{\mu} k^{\mathcal{P}}_{\rho} k^{\mathcal{P}}_{\sigma} \sum_{ \mathcal{P} = \mathcal{Q} \cup \mathcal{R} } \bigg( J^{\mathcal{Q}}_{\rho\sigma} J^{\mathcal{R}}_{\mu\nu} 
  - J^{\mathcal{Q}}_{\mu\rho} J^{\mathcal{R}}_{\sigma\nu} \bigg)\,,
  \\
  &= - \frac{1}{s^{\mathcal{P}}} k^{\mathcal{P}}_{\mu} k^{\mathcal{P}}_{\rho} k^{\mathcal{P}}_{\sigma} \sum_{ \mathcal{P} = \mathcal{Q} \cup \mathcal{R} } \bigg( J^{\mathcal{Q}}_{\rho\sigma} J^{\mathcal{R}}_{\mu\nu} 
  - J^{\mathcal{Q}}_{\rho\sigma} J^{\mathcal{R}}_{\mu\nu} \bigg) =0\,.
\end{aligned}\label{}
\end{equation}
%

\section{Conclusion}\label{Sec:6}
This paper presents the classical double copy for graviton off-shell currents. To this end, we have first formulated the perturbative DFT with all-orders of perturbations of the generalised metric. We have found the explicit field redefinition between the perturbation of the generalised metric and metric perturbations in GR. This shows the equivalence between perturbative DFT and the perturbative GR. The key point of this paper was to show that the graviton off-shell currents satisfy the current KLT relation representing the graviton currents in terms of the gluon currents. 

Next we have constructed the off-shell recursion relation for gravity from the perturbiner method for the perturbative DFT instead of the recursive structure of the Feynman vertices. We obtained the graviton off-shell currents up to rank-4 by solving the graviton off-shell recursion relation iteratively. This method directly connects the graviton scattering amplitudes to the solution of the perturbative EoM. Thus it fits with our purpose of understanding the perturbative classical double copy via the current KLT relation. Using the explicit graviton currents, we have presented the current KLT relation that relates the gluon and the graviton currents up to the gauge transformations and the regular terms. As a nontrivial test for the graviton off-shell currents, we checked the off-shell conservation, a characteristic of the currents. However, we showed that the graviton current conservation does not hold exactly for our nonlinear gage conditions, but the currents are conserved up to the gauge condition. To support this, we examined the gluon currents with the Gervais-Neveu gauge condition, which is nonlinear in the gauge field, are conserved up to the gauge condition.

In this work, we have focused on the pure gravity case for simplicity. Since DFT naturally unifies the other massless NSNS fields, the Kalb-Ramond field $B_{\mu\nu}$ and dilaton $\phi$, as well as the gauge multiplet in the heterotic string, it would be straightforward to apply the same procedure in pure gravity to the entire NSNS sector or $N=4$ supergravities. Another essential feature of DFT that we have left out is the doubled local Lorentz group. It should be related to the manifest double copy structure for the graviton currents, and it would be helpful for describing the double copy structure in various applications of perturbative gravity. We remain these as future work.

An important issue related to the classical double copy is how to construct the exact current KLT relation for general helicities by removing the regular terms and the gauge transformation. Once we have the exact KLT relation, we can directly write down perturbative gravity solutions using gauge theory solutions, as in the scattering amplitude. From the perturbiner method point of view, this issue is related to the form of EoM because the field equations determine the form of off-shell recursions and currents. Note that EoM is not unique since we can freely add or subtract trivial terms that vanish in on-shell (in a sense that satisfies the EoM). Also, gauge choice and field redefinitions change the form of EoM. We will investigate the exact KLT relation in future work.

It is also interesting to extend our results to curved backgrounds. It would be straightforward to apply the flat background results to curved backgrounds. As we have pointed out, the perturbative DFT is simpler than GR due to the $\mathit{O}(D,D)$ structure and doubled Lorentz group. We can apply the perturbative DFT to various perturbative gravity applications, particularly black holes or cosmological backgrounds, especially (A)dS space. Since the perturbiner methods are not restricted to a flat background, we may apply them to understand the (classical) double copy for curved backgrounds.

\acknowledgments
We thank Miok Park for helping us to use Mathematica code computing the recursion relation. This work is supported by an appointment to the JRG Program at the APCTP through the Science and Technology Promotion Fund and Lottery Fund of the Korean Government. It is also supported by the National Research Foundation of Korea(NRF) grant funded by the Korea government(MSIT) No.2021R1F1A1060947 and the Korean Local Governments of Gyeongsangbuk-do Province and Pohang City.

\newpage
\appendix

\section{Explicit form of the Graviton Recursions}\label{appendix_A}
In this appendix we present the explicit expressions of the graviton recursion relation by expanding the currents of the graviton and the auxiliary fields \eqref{graviton_recursion_relation} and \eqref{auxiliary_recursion_relation} up to rank-4. 
\begin{itemize}
\item \textbf{Rank-1}
\\
In this case the graviton currents are identified with the polarization tensor $J_{\mu\nu}^{i} = \epsilon^{i}_{\mu\nu}$ as the initial condition for the recursion. The auxiliary currents vanish $\varDelta_{\mu\nu}^{i} = \mathcal{X}^{i}_{\mu\nu} = \mathcal{Y}^{i}_{\mu\nu} = \mathcal{Z}^{i}_{\mu\nu} =\mathcal{W}^{i}_{\mu\nu}=0\,.$ Then the rank-1 recursion gives the mass-shell condition $(k^i)^2 =0$. 
\begin{equation}
  \big(k^{i}\cdot k^{i}\big) \mathcal{J}_{\mu\nu}^{i} = \big(k^{i}\big)^{2} \epsilon^{i}_{\mu\nu}= 0\,.
\label{}\end{equation}
\item \textbf{Rank-2}
\\
The rank-2 graviton currents are given by
\begin{equation}
\begin{aligned}
  \mathcal{J}^{ij}_{\mu\nu} &= \frac{1}{s_{ij}} \big(\mathcal{X}^{ij} +\mathcal{Y}^{ij}+\mathcal{Z}^{ij}\big)_{\mu\nu}\,,
\end{aligned}\label{rank2_1}
\end{equation}
and the auxiliary currents are
\begin{equation}
\begin{aligned}
  \Delta_{\mu \nu}^{i j}&=\frac{1}{2} J_{\mu \rho}^{i} J_{\nu \rho}^{j} + (i\leftrightarrow j)\,,
  \\
  \mathcal{X}^{ij}_{\mu\nu} &= \frac{1}{2} k^{i}_{(\mu} k^{j}_{\nu)} \mathcal{J}^{i}_{\rho\sigma} \mathcal{J}^{j}_{\rho\sigma}
  -2 k_{(\mu}^{i} \mathcal{J}^{j}_{\nu)\rho} \mathcal{J}^{i}_{\rho\sigma} k^{j}_{\sigma}
  + (i\leftrightarrow j)\,,
  \\
  \mathcal{Y}^{ij}_{\mu\nu} &= (\mathcal{J}^{i}_{\mu\rho} k^{j}_{\rho}\big) \big(\mathcal{J}^{j}_{\nu\sigma}k^{i}_{\sigma}\big)
  + (i\leftrightarrow j)\,,
  \\
  \mathcal{Z}^{ij}_{\mu\nu} &= - \big(k^{j}_{\rho} k^{j}_{\sigma} \mathcal{J}^{i}_{\rho\sigma}\big) \mathcal{J}^{j}_{\mu\nu}
  + (i\leftrightarrow j)\,,
  \\
  \mathcal{W}^{ij}_{\mu\nu} &= s_{ij} \varDelta^{ij}_{\mu\nu}  + (i\leftrightarrow j)\,.
\end{aligned}\label{rank2_2}
\end{equation}
As we have assumed, all the currents are symmetric under the exchanging $i$ and $j$. 
\item \textbf{Rank-3}\\
The rank-3 graviton currents are represented by
\begin{equation}
\begin{aligned}
  \mathcal{J}^{ijk}_{\mu\nu} &= \frac{1}{s_{ijk}}\big(\mathcal{X}^{ijk}_{\mu\nu}+\mathcal{Y}^{ijk}_{\mu\nu}+\mathcal{Z}^{ijk}_{\mu\nu}\big) +\frac{1}{2!}\frac{1}{s_{ijk}} \sum_{\text{Perm}[ijk]}
  \mathcal{J}^{i}_{\rho(\mu} \big(- \mathcal{X}^{jk} +\mathcal{Y}^{jk}+\mathcal{W}^{jk} \big)_{\nu)\rho}\,,
\end{aligned}\label{rank3_graviton_current}
\end{equation}
and the auxiliary fields are
\begin{equation}
\begin{aligned}
  \varDelta^{ijk}_{\mu \nu} &= \frac{1}{4} \sum_{\text{Perm}[ijk]} \mathcal{J}^{i}_{\mu \rho} \mathcal{J}^{jk}_{\rho \nu}\,,
  \\
  \mathcal{X}^{ijk}_{\mu\nu} &= \sum_{\text{Perm}[ijk]} \bigg[\
    \frac{1}{4} k^{i}_{\mu} k^{jk}_{\nu} \mathcal{J}^{i}_{\rho \sigma} ( \mathcal{J}^{jk}_{\rho \sigma} + \varDelta^{jk}_{\rho \sigma} )
  + \frac{1}{4} k^{ij}_{\mu} k^{k}_{\nu} (\mathcal{J}^{ij}_{\rho \sigma}  - \varDelta^{ij}_{\rho \sigma}) \mathcal{J}^{k}_{\rho \sigma}\,, 
  \\
  &  \qquad\qquad\qquad
  - k^{i}_{(\mu}  \left( \mathcal{J}+ \varDelta  \right)^{jk}_{\nu) \rho} \mathcal{J}^{i}{}^{\rho \sigma} {k}^{jk}_{\sigma}
  - k^{ij}_{(\mu} \mathcal{J}^{k}_{\nu) \rho} (\mathcal{J}^{ij}_{\rho \sigma} + \varDelta^{ij}_{\rho \sigma}) {k}^{k}_{\sigma} \bigg]\,,
  \\
   \mathcal{Y}^{ijk}_{\mu\nu} & = \sum_{\text{Perm}[ijk]} \bigg[\
    \frac{1}{2} (\mathcal{J}^{i}_{\mu\rho} k^{jk}_{\rho}\big) \big(\mathcal{J}^{jk}_{\nu\sigma}-\Delta^{jk}_{\nu\sigma}\big) k^{i}_{\sigma}
  + (\mu\leftrightarrow \nu) \ \bigg]\,,
  \\
  \mathcal{Z}^{ijk}_{\mu\nu} &= - \frac{1}{2!}\sum_{\text{Perm}[ijk]} \bigg[
  \big(k^{jk}_{\rho} k^{jk}_{\sigma} \mathcal{J}^{i}_{\rho\sigma}\big) \mathcal{J}^{jk}_{\mu\nu}
  + k^{k}_{\rho} k^{k}_{\sigma}
  \big(\mathcal{J}^{ij}_{\rho\sigma}-\varDelta^{ij}_{\rho\sigma}\big) \mathcal{J}^{k}_{\mu\nu}\bigg]\,,
  \\
  \mathcal{W}^{ijk}_{\mu\nu}&= s_{ijk} \varDelta_{\mu\nu}^{ijk}
  + \frac{1}{2!}\sum_{\text{Perm}[ijk]} \big(k^{jk}_{\rho} k^{jk}_{\sigma}\mathcal{J}^{i}_{\rho\sigma}\big) \varDelta^{jk}_{\mu\nu}\,.
\end{aligned}\label{}
\end{equation}
Here $\displaystyle \sum_{\text{Perm}[i,j,k]}$ denotes the sum over all permutations of the set $\{i,j,k\}$. It can be extended to the higher permutation group straightforwardly. 
~\\~\\
\item \textbf{Rank-4}
\\
The rank-4 graviton current is given by
\begin{equation}
\begin{aligned}
    \mathcal{J}^{ijkl}_{\mu\nu} &= \frac{1}{s_{ijkl}}\big(\mathcal{X}^{ijkl}_{\mu\nu}+\mathcal{Y}^{ijkl}_{\mu\nu}+\mathcal{Z}^{ijkl}_{\mu\nu}\big) 
  \\
  &+ \frac{1}{s_{ijk}} \sum_{\text{Perm}[ijkl]} \bigg[\
  \frac{1}{3!} \mathcal{J}^{i}_{\rho(\mu} \big(- \mathcal{X}^{jkl} +\mathcal{Y}^{jkl}+\mathcal{W}^{jkl} \big)_{\nu)\rho}
  + \frac{1}{4}  \mathcal{J}^{ij}_{\rho(\mu} \big(- \mathcal{X}^{kl} +\mathcal{Y}^{kl}+\mathcal{W}^{kl} \big)_{\nu)\rho}
  \\
  & +\frac{1}{4} \varDelta^{ij}_{\rho(\mu} \big(\mathcal{X}^{kl}+\mathcal{Y}^{kl} +\mathcal{Z}^{kl} - s_{kl} \varDelta^{kl} \big)_{\nu)\rho}\  \bigg]\,,
\end{aligned}\label{}
\end{equation}
and the auxiliary currents are
\begin{equation}
\begin{aligned}
    \varDelta^{ijkl}_{\mu \nu} &= \frac{1}{2} \sum_{\text{Perm}[ijkl]}  \bigg[ \frac{1}{3!}\mathcal{J}^{i}_{\mu \rho} \mathcal{J}^{jkl}_{\rho \nu}
  + \frac{1}{4} \mathcal{J}^{ij}_{\mu \rho} \mathcal{J}^{kl}_{\rho \nu}  
  - \frac{1}{16} (\mathcal{J})^{2 \phantom{1} ij}_{\mu \rho} (\mathcal{J})^{2 \phantom{1} kl}_{\nu \rho}\bigg]\,,
  \\ 
  \mathcal{X}^{ijkl}_{\mu\nu} &= \frac{1}{2} \sum_{\text{Perm}[ijkl]} \bigg[\
    \frac{1}{3!} k^{i}_{(\mu} k^{jkl}_{\nu)}  \mathcal{J}^{i}_{\rho \sigma} (\mathcal{J} + \varDelta)^{jkl}_{\rho \sigma}
  + \frac{1}{4} k^{ij}_{(\mu} k^{kl}_{\nu)}  (\mathcal{J} - \varDelta)^{ij}_{\rho \sigma} (\mathcal{J} + \varDelta)^{kl}_{\rho \sigma}
  \\
  &\qquad\qquad\qquad\quad + \frac{1}{3!} k^{ijk}_{(\mu} k^{l}_{\nu)}  (\mathcal{J} - \varDelta)^{ijk}_{\rho \sigma} \mathcal{J}^{l}_{\rho \sigma}\ \bigg]\,,
  \\
  &\quad  -2 \sum_{\text{Perm}[ijkl]} \bigg[\ \frac{1}{3!} k^{i}_{(\mu} (\mathcal{J} + \varDelta)^{jkl}_{\nu) \rho} \mathcal{J}^{i}_{\rho \sigma} {k}^{jkl}_{\sigma}
  + \frac{1}{4} k^{ij}_{(\mu} (\mathcal{J} + \varDelta)^{kl}_{\nu) \rho} (\mathcal{J} - \varDelta)^{ij}_{\rho \sigma}{k}^{kl}_{\sigma}
  \\
  &\qquad\qquad\qquad\quad + \frac{1}{3!} k^{ijk}_{(\mu} \mathcal{J}^{l}_{\nu) \rho} (\mathcal{J} - \varDelta)^{ijk}_{\rho \sigma} {k}^{l}_{\sigma}\ \bigg]\,,
  \\
  \mathcal{Y}^{ijkl}_{\mu\nu} &= \sum_{\text{Perm}[ijkl]} \bigg[\
    \frac{1}{3!} (\mathcal{J} - \varDelta)^{ijk}_{\mu \rho} {k}^{l}_{\rho} \mathcal{J}^{l}_{\nu \sigma} {k}^{ijk}_{\sigma}
  + \frac{1}{4} (\mathcal{J}- \varDelta)^{ij}_{\mu \rho}  {k}^{kl}_{\rho}(\mathcal{J} - \varDelta)^{kl}_{\nu \sigma} {k}^{ij}_{\sigma}
  \\
  &\qquad\qquad\qquad + \frac{1}{3!} \mathcal{J}^{i}_{\mu \rho} {k}^{jkl}_{\rho} (\mathcal{J} - \varDelta)^{jkl}_{\nu \sigma} {k}^{i}_{\sigma}\ \bigg]\,, 
  \\
  \mathcal{Z}^{ijkl}_{\mu\nu} & = -\sum_{\text{Perm}[ijkl]} \bigg[\
   \frac{1}{4} k^{ij}_{\rho} k^{ij}_{\sigma} (J^{kl}_{\rho \sigma} - \varDelta^{kl}_{\rho \sigma}) \varDelta^{ij}_{\mu \nu}
   + \frac{1}{3!} k^{ijk}_{\rho} k^{ijk}_{\sigma} (J^{i}_{\rho \sigma} - \varDelta^{i}_{\rho \sigma}) \varDelta^{ijk}_{\mu \nu} \bigg]\,, 
  \\
  \mathcal{W}^{ijkl}_{\mu\nu} &=  \sum_{\text{Perm}[ijkl]} \bigg[\ \frac{1}{4} k^{ij}_{\rho} k^{ij}_{\sigma} ( \mathcal{J}^{kl}_{\rho \sigma} - \varDelta^{kl}_{\rho \sigma}) \varDelta^{ij}_{\mu \nu}
  + \frac{1}{3!} k^{ijk}_{\rho} k^{ijk}_{\sigma} ( \mathcal{J}^{l}_{\rho \sigma} - \varDelta^{l}_{\rho \sigma}) \varDelta^{ijk}_{\mu \nu} \bigg]
  \\
  &\quad +s_{ijkl} \varDelta^{ijkl}_{\mu \nu} \,.
\end{aligned}\label{}
\end{equation}
\end{itemize}

\section{Gluon off-shell currents in the Lorentz gauge}\label{appendix_B}

We present the explicit form of the gluon currents with up to rank-4 by solving the gluon recursion relations in \eqref{YM_recursions}.
\\~\\
\textbf{$\bullet$ Rank-1}
\\
The rank-1 current is the initial condition of the BG recursion relation, which is identified with the polarization vectors
\begin{equation}
  J_{\mu}^{i} = \epsilon^{i}_{\mu}
\label{}\end{equation}
The rank-1 field strength is given by
\begin{equation}
  F^{i}_{\mu\nu} = i k_{\mu}^{i} J_{\nu}^{i} -i k_{\nu}^{i} J_{\mu}^{i}\,.
\label{}\end{equation}
\\
\textbf{$\bullet$ Rank-2}
\\
The rank-2 recursion relation is given by
\begin{equation}
\begin{aligned}
  J^{ij}_{\mu} &= \frac{i}{\sqrt{2} s^{ij}} \Big(i\left(J^{i} \cdot k^{j}\right) J_{\mu}^{j}+J_{\nu}^{i} F_{\nu \mu}^{j}-(i \leftrightarrow j)\Big) 
  \\
  F^{ij}_{\mu\nu} &= i \Big(k^{i j}_{\mu} J^{i j}_{\nu} - k^{i j}_{\nu} J^{i j}_{\mu} - \frac{1}{\sqrt{2}} \big(J_{\mu}^{i}J_{\nu}^{j} -J_{\nu}^{i}J_{\mu}^{j} \big)\Big)
\end{aligned}\label{}
\end{equation}
If we substitute the initial condition into the rank-2 current, we get easily the $J^{ij}_{\mu}$ as
\begin{equation}
\begin{aligned}
  J^{ij}_{\mu} &= \frac{\epsilon_{i}{\cdot}\epsilon_{j} \left(k^j_{\mu} -k^i_{\mu}\right)+2\big(k_{i}{\cdot}\epsilon_{j} \big)\epsilon^i_{\mu} -2\big(k_{j}{\cdot}\epsilon_{i}\big) \epsilon^j_{\mu}}{\sqrt{2} s_{ij}}\,.
\end{aligned}\label{}
\end{equation}
\\
\textbf{$\bullet$ Rank-3}  
\\
The recursion relation for the rank-3 current is given by
\begin{equation}
\begin{aligned}
  J^{ijk}_{\mu} &= \frac{i}{\sqrt{2}s_{ijk}}\Big( i \big(J^{i}\cdot k^{jk}\big)J^{jk}_{\mu}+J^{i}_{\nu}F^{jk}_{\nu\mu} - \big(i\leftrightarrow jk\big)
  \\
  &\qquad\qquad\quad +i \big(J^{ij}\cdot k^{k}\big)J^{k}_{\mu} +J^{ij}_{\nu}F^{k}_{\nu\mu} - (ij\leftrightarrow k)
  \Big)\,,
    \\
  F^{ijk}_{\mu\nu} &= 2i k^{ijk}_{[\mu} J^{ijk}_{\nu]} - i\sqrt{2}\Big( J^{i}_{[\mu} J^{jk}_{\nu]} + J^{ij}_{[\mu} J^{k}_{\nu]} \Big)\,.
\end{aligned}\label{}
\end{equation}
We can solve the recursion relation by substituting the rank-1 and rank-2 currents
\begin{equation}
\begin{aligned}
  J_{\mu}^{ijk} &= \frac{2}{s_{ijk}} \bigg[\ 
  \frac{ Q_{ikj} k^j_{\mu} - Q_{jki} k^{i}_{\mu} 
  + R_{ijk} k^k_{\mu}}{s_{12}} 
  +\frac{ Q_{kij}k^j_{\mu} -Q_{jik} k^k_{\mu} 
  +R_{kji} k^i_{\mu}}{s_{23}}
  \\
  &\qquad\qquad -\left(
  \frac{s_{ik}\big(P_{ijk}+P_{ikj}\big)}{s_{jk}}
  + \frac{s_{ik}(\epsilon_{j}{\cdot}\epsilon_{k})}{2s_{jk}}\right) \epsilon^1_{\mu}
  -\left(P_{jik} +P_{jki}+\frac{\epsilon_{i}{\cdot}\epsilon_{k}}{j} \right) \epsilon^j_{\mu}
  \\
  &\qquad\qquad -\left(\frac{s_{ik}\big(P_{kji}+P_{kij}\big)}{s_{ij}}
   +\frac{s_{ik} (\epsilon_{i}{\cdot}\epsilon_{j})}{2s_{ij}}\right) \epsilon^k_{\mu} \ \bigg] + \text{regular terms} + k^{ijk}_{\mu} \Psi^{ijk} \,,
\end{aligned}\label{J123}
\end{equation}
where
\begin{equation}
  \text{regular terms} :~  \bigg(2P_{ijk}+\frac{\epsilon_{j}{\cdot}\epsilon_{k}}{2}\bigg)\frac{\epsilon^i_{\mu }}{s_{jk}}
  +\Big(2P_{kji}+\frac{\epsilon_{i}{\cdot}\epsilon_{j}}{2} \Big)\frac{\epsilon^{k}_{\mu}}{s_{12}}\,.
\label{}\end{equation}
The last term in \eqref{J123} can be interpreted to a gauge transformation and does not contribute to the scattering amplitude
\begin{equation}
\begin{aligned}
  \Psi^{123} &= \frac{1}{2 s_{ijk}} \bigg[\frac{1}{s_{12}} \Big(Q_{231}-Q_{132}-2R_{123} \Big)
- \frac{1}{s_{23}} \Big(Q_{312} -Q_{213}-2R_{231} \Big)\bigg]\,.
\end{aligned}\label{}
\end{equation}
\\
\textbf{$\bullet$ Rank-4}
\\
For the rank-4 current KLT relation, we need six gluon currents: $J^{1234}_{\mu}$, $J^{1243}_{\mu}$, $J^{1324}_{\mu}$, $J^{1342}_{\mu}$, $J^{1423}_{\mu}$ and $J^{1432}_{\mu}$. We introduce the following functions generalising \eqref{PQR} 
\begin{equation}
\begin{aligned}
    &P_{ijk} = \frac{ k_{i}{\cdot}\epsilon_{j} ( k_{i}{\cdot}\epsilon_{k} + k_{j}{\cdot}\epsilon_{k})}{s_{ij}}, \qquad Q_{ijk} = (k_{i}{\cdot}\epsilon_{j}) (\epsilon_{i}{\cdot}\epsilon_{k}) , \qquad R_{ijk} = Q_{ijk} - Q_{jik}\,, \\
  & S^{\pm}_{ijk} = Q_{ijk} \pm Q_{kji}\,, \qquad T_{ijk} = R_{ijk} + Q_{jki}\,, \\
  & Q_{ijk,l} = Q_{ijk} (k_{k}{\cdot}\epsilon_{l}), \qquad Q_{i,jkl} = (k_{i}{\cdot}\epsilon_{j}) Q_{jkl}\,, \qquad
  Q'_{i,jkl} = (k_{j}{\cdot}\epsilon_{i}) Q_{jkl}\,,  \\
  & P_{i,jkl} = (k_{i}{\cdot}\epsilon_{j}) P_{jkl}\,, \qquad P'_{i,jkl} = (k_{j}{\cdot}\epsilon_{i}) P_{jkl}, \qquad P_{ijk,l} = P_{ijk} (k_{j}{\cdot}\epsilon_{l} + k_{k}{\cdot}\epsilon_{l})\,.
\end{aligned}\label{}
\end{equation}
We introduce a rescaled currents $\hat{J}_{\mu}^{P} = \frac{s_{1234}}{2\sqrt{2}} J^{P}_{\mu}$ for simplicity. Then the explicit form of the rank-4 gluon currents are as follows:
\\
$1.~ \displaystyle \frac{s_{1234}}{2\sqrt{2}}  J^{1234}_{\mu}$
\begin{equation}
\begin{aligned}
  \hat{J}^{1234}_{\mu} &= - \bigg[ \frac{1}{s_{23}}- \frac{s_{13}+s_{14}}{s_{12} s_{34}} \bigg] \frac{Q_{214,3}}{s_{234}} k^{1}_{\mu}
- \frac{1}{s_{34}}\bigg[ \frac{Q_{214,3}}{s_{12}} + \frac{ Q_{214,3} + Q'_{1,432} }{ s_{234} } \bigg] k^{2}_{\mu}
\\
& \quad
- \frac{1}{s_{23}} \bigg[ \frac{(s_{12}+s_{23})Q'_{1,234}}{s_{12} s_{123}} + \frac{Q'_{1,234} + Q_{234,1}}{  s_{234} } \bigg] k^{4}_{\mu}
\\
& \quad - \bigg[ \frac{(s_{13}+s_{14})\big(P_{123} (k_{3} {\cdot} \epsilon_{4}) - P_{124} (k_{4} {\cdot} \epsilon_{3})\big)}{s_{34} s_{234}} 
- \frac{(s_{12}+s_{23} )P_{1,234}}{s_{12}s_{123}} - \frac{P_{1,234}}{s_{234}} 
\\
& \qquad \quad
-  \frac{( s_{13} + s_{14})Q_{432} }{2 s_{34} s_{234}} 
+ \frac{s_{14}Q_{234}}{2s_{23}s_{234}} \bigg] \epsilon^{1}_{\mu}
\\
& \quad
  {+} \bigg[  \frac{(s_{13}{+}s_{14})(P_{213,4} {-}P_{214,3})}{s_{34} s_{234}}  -  \frac{(s_{12}{+}s_{23})P'_{1,234}}{s_{12} s_{123}} 
- \frac{P'_{1,234} {+} P_{234,1} {+} P_{2,341} {-} P_{2,431}}{s_{234}} \bigg] \epsilon^{2}_{\mu}
\\
& \quad
+ \bigg[ \frac{ P_{342} (k_{2} {\cdot} \epsilon_{1}) - P_{341} (k_{1} {\cdot} \epsilon_{2}) }{s_{12}} + \frac{s_{24} P_{3,421} }{ s_{34} s_{234} } + \frac{Q_{214}}{2 s_{12}} + \frac{S^{+}_{214}}{2 s_{234}} \bigg] \epsilon^{3}_{\mu}
\\
& \quad  {+} \bigg[  \frac{(s_{12} + s_{23})(P_{4,213} - P_{4,123} )}{s_{23}s_{123}} + \frac{ P_{431,2} - P_{432,1} }{ s_{12} } + \frac{s_{24}P_{421,3}}{s_{23}s_{234}}
  {-} \frac{s_{24} P'_{3,421}}{ s_{34} s_{234}}  \bigg]\epsilon^{4}_{\mu}
\\
& \quad  +s_{1234} \big(\text{regular terms}\big) + k^{1234}_{\mu} \Lambda^{1234}
\end{aligned}
\end{equation}
where the regular terms are
\begin{equation}
  - \frac{Q_{214,3}}{s_{12} s_{34} s_{234}} k^{1}_{\mu} 
  + \frac{1}{s_{234}}\bigg[  \frac{ ( P_{123}k_{3} {\cdot} \epsilon_{4} - P_{124}k_{4} {\cdot} \epsilon_{3}   )}{s_{34}}   +  \frac{Q_{234}}{4 s_{23}} -\frac{Q_{432}}{4 s_{34}} \bigg] \epsilon^{1}_{\mu}
  + \frac{P_{214,3}-P_{213,4}}{s_{34}s_{234}}   \epsilon^{2}_{\mu}\,,
\label{}\end{equation}
and
\begin{equation}
\Lambda^{1234}=\frac{1}{2} \bigg[ \frac{Q_{214,3}}{s_{12}s_{34}} 
+ \frac{(s_{12}+s_{23}) Q'_{1,234}}{ s_{12}s_{23}s_{123}} + \frac{3Q_{214,3}+Q'_{1,234}}{2 s_{23} s_{234}}
+ \frac{3Q_{214,3}+Q'_{1,432}}{2 s_{34} s_{234}} \bigg]\,.
\end{equation}
\\
${ 2.~ \displaystyle \frac{s_{1234}}{2\sqrt{2}}  J^{1243}_{\mu}}$
\begin{equation}
\begin{aligned}
  \hat{J}^{1243}_{\mu} &=\bigg[ \frac{1}{ s_{234} }-\frac{s_{13}+s_{23}}{s_{12}s_{124}} \bigg] \frac{Q_{214,3}}{s_{34}} k^{1}_{\mu} - \bigg[ \frac{Q'_{1,234}}{s_{12} s_{124}} - \frac{(s_{13}-s_{24})(Q_{234,1}+Q'_{1,234})}{s_{24}s_{124}s_{234}}  \bigg] k^{4}_{\mu}
  \\
  & \quad
  - \bigg[ \frac{(s_{13}+s_{23})Q_{214,3}}{s_{12}s_{34}s_{124}} 
  - \frac{ (s_{13} s_{24} + s_{13} s_{34} + s_{23}s_{24}) (Q_{214,3} + Q'_{1,432})}{s_{24}s_{34}s_{124}s_{234}} \bigg] k^{2}_{\mu}
  \\
  & \quad - \bigg[ \frac{(s_{13}+s_{14}) \big(P_{124} (k_{4} {\cdot} \epsilon_{3}) - P_{123}(k_{3} {\cdot} \epsilon_{4})\big)}{s_{34}s_{234}} 
  + \frac{(s_{13} - s_{24} )P_{1,243}}{s_{124}s_{234}}
  \\
  &\qquad \quad - \frac{s_{24} P_{1,243}}{s_{12} s_{124}}
  - \frac{s_{14} (s_{13}-s_{24})S^{+}_{234}}{2 s_{24} s_{124}s_{234}}
  +\bigg( \frac{s_{13}+s_{14}}{s_{34}}+\frac{s_{13}}{s_{24}} \bigg) \frac{Q_{432}}{2s_{234}}
  \bigg] \epsilon^{1}_{\mu}
  \\
  & \quad
  + \bigg[ \frac{(s_{13} {+} s_{14})(P_{214,3} {-} P_{213,4})}{s_{34}s_{234}}  + \frac{(s_{13}{-}s_{24})(P'_{1,243} {+} P_{243,1})}{s_{124}s_{234}} 
  - \frac{s_{24} P'_{1,243}}{s_{12} s_{124}}
  +\frac{ P_{2,341} {-} P_{2,431} }{s_{234}} \bigg] \epsilon^{2}_{\mu}
  \\
  & \quad + \bigg[ \frac{(s_{13}+s_{23})\big(P_{342}(k_{2} {\cdot} \epsilon_{1})  - P_{341}(k_{1} {\cdot} \epsilon_{2})\big)}{s_{12}s_{124}}  
- \frac{s_{23}(s_{13}-s_{24})P_{3,421}}{s_{34}s_{124}s_{234}}
\\
& \qquad \quad
+\frac{s_{13}P_{3,421}}{s_{34} s_{124} } 
- \frac{ s_{23} ( s_{13} - s_{24}) S^{+}_{214} }{2 s_{24} s_{124} s_{234}}+ \bigg( \frac{s_{13}+s_{23}}{s_{12}} + \frac{s_{13}}{s_{24}} \bigg) \frac{Q_{214}}{2s_{124}}  \bigg] \epsilon^{3}_{\mu}
\\
& \quad
- \bigg[ \frac{(s_{13}+s_{23})( P_{432,1} - P_{431,2})}{s_{12} s_{124}}
 + \frac{(s_{13}-s_{24})(P'_{3,421}+P_{421,3}) }{s_{124}s_{234}}
\\
& \qquad \quad
- \frac{s_{24}P'_{3,421}}{s_{34} s_{234}} 
+ \frac{P_{4,123}-P_{4,213}}{s_{124}}  \bigg] \epsilon^{4}_{\mu} + s_{1234} \big(\text{regular terms}\big) + k^{1234}_{\mu} \Lambda^{1243}
\end{aligned}
\end{equation}
where the regular terms are
\begin{equation}
\begin{aligned}
  &\frac{Q_{214,3}}{s_{12} s_{34} s_{124}} k^{1}_{\mu}
+\frac{1}{s_{34} s_{124}}\bigg[ \frac{ Q_{214,3}}{s_{12}} 
+\frac{(s_{24} + s_{34})(Q_{214,3}+Q'_{1,432})}{s_{24}s_{234}} \bigg] k^{2}_{\mu} 
+\frac{Q_{214,3} + Q'_{1,234}}{s_{24} s_{124}s_{234}} k^{4}_{\mu}
\\
&\quad
+ \frac{1}{s_{234}}\bigg[ \frac{(P_{124} k_{4} {\cdot} \epsilon_{3} - P_{123}k_{3} {\cdot} \epsilon_{4})}{s_{34}} 
+ \frac{P_{1,243}}{s_{124} }
- \frac{s_{14}  S^{+}_{234}}{2 s_{24} s_{124}}
+ \frac{(Q_{234} + 3 Q_{432})}{8 s_{24} }
+\frac{Q_{432}}{4s_{34}}
\bigg] \epsilon^{1}_{\mu}
\\
  & - \frac{1}{s_{234}}\bigg[ \frac{P_{214,3} - P_{213,4}}{s_{34} } 
  + \frac{P'_{1,243} + P_{243,1} }{s_{124}} \bigg] \epsilon^{2}_{\mu}
  \\
  & -\frac{1}{s_{124}} \bigg[ \frac{(P_{342}k_{2} {\cdot} \epsilon_{1}  - P_{341}k_{1} {\cdot} \epsilon_{2})}{s_{12}}  
  +\frac{(s_{24} + s_{34})  P_{3,421}}{s_{34}  s_{234}} 
  - \frac{ s_{23}  S^{+}_{214} }{2 s_{24}  s_{234}}
  {+}\frac{(Q_{412}{+}3Q_{214})}{8s_{24}}
  {+}\frac{Q_{214}}{4 s_{12}} \bigg] \epsilon^{3}_{\mu}
  \\
  & + \frac{1}{s_{124}}\bigg[ \frac{ P_{432,1} - P_{431,2}}{s_{12} }
+ \frac{P'_{3,421}+P_{421,3} }{s_{234}}\bigg] \epsilon^{4}_{\mu}
- \frac{(Q'_{1,432}-Q'_{1,234})}{8s_{24} s_{124} s_{234}} k^{1234}_{\mu}\,.
\end{aligned}\label{}
\end{equation}
and
\begin{equation}
\begin{aligned}
  \Lambda^{1243} = \frac{1}{2} \bigg[ \frac{(s_{13}{-}s_{24})(Q'_{1,432} {-}Q'_{1,234})}{4s_{24} s_{124} s_{234}} - \frac{Q_{214,3}}{s_{12}s_{34}} - \frac{Q_{214,3}{-}Q'_{1,234}}{2s_{12}s_{124}} - \frac{3Q_{214,3}{+}Q'_{1,432}}{2s_{34}s_{234}} \bigg] 
\end{aligned}
\end{equation}
\\~\\
$3.~ \displaystyle \frac{s_{1234}}{2\sqrt{2}}  J^{1324}_{\mu}$
\begin{equation}
\begin{aligned}
  \hat{J}^{1324}_{\mu} &= \frac{Q_{234,1}}{s_{23} s_{234}} k^{1}_{\mu} - \frac{Q_{214,3}{+}Q'_{1,432}}{s_{24} s_{234} }k^{2}_{\mu}
  + \frac{1}{s_{23}} \bigg[ \frac{Q'_{1,234}}{s_{123}} {+}  \frac{(s_{23}{+}s_{24})(Q_{234,1}{+}Q'_{1,234})}{ s_{24} s_{234}} \bigg] k^{4}_{\mu}
  \\
  & \quad - \bigg[  \frac{P_{1,234}+P_{1,243}}{s_{234}} 
  +  \frac{P_{1,234}}{s_{123}} 
  + \frac{s_{12} Q_{432} -s_{14} Q_{234} }{2 s_{24} s_{234}}
  - \frac{s_{14} Q_{234} }{2 s_{23} s_{234}}
  \bigg] \epsilon^{1}_{\mu}
  \\
  & \quad
  + \bigg[ \frac{P'_{1,234}}{s_{123}} + \frac{ P'_{1,234}+P'_{1,243}+P_{234,1} + P_{243,1}}{s_{234}} \bigg] \epsilon^{2}_{\mu}
  \\
  & \quad
  + \bigg[ \frac{P_{3,421}}{s_{234}} 
  +\frac{s_{34} Q_{214} - (s_{23}+s_{24})Q_{412}}{2s_{24} s_{234}} \bigg]\epsilon^{3}_{\mu}
  \\
  & \quad
  - \bigg[ \frac{s_{12}(P_{4,213}-P_{4,123}) }{s_{23} s_{123}} + \frac{ s_{24} P_{421,3}  }{s_{23} s_{234}} + \frac{P_{421,3}+P'_{3,421} }{s_{234}} \bigg] \epsilon^{4}_{\mu}
  \\
  &\quad + s_{1234} \big(\text{regular terms}\big) + k^{1234}_{\mu} \Lambda^{1324}
\end{aligned}
\end{equation}
where the regular terms are
\begin{equation}
\begin{aligned}
  & -\frac{1}{4 s_{234}}\bigg[ \frac{Q_{234}-Q_{432}}{2 s_{24}} + \frac{Q_{234}}{s_{23}}\bigg] \epsilon^{1}_{\mu}
\end{aligned}\label{}
\end{equation}
and
\begin{equation}
\begin{aligned}
\Lambda^{1324}=- \frac{1}{2} \bigg[ \frac{Q'_{1,234}}{s_{23} s_{123}} 
 +\frac{Q'_{1,234}-Q'_{1,432}}{4s_{24} s_{234}}
+ \frac{Q'_{1,234}+3Q_{214,3}}{2s_{23} s_{234}} \bigg]
\end{aligned}
\end{equation}
\\~\\
$4.~ \displaystyle \frac{s_{1234}}{2\sqrt{2}}  J^{1342}_{\mu}$
\begin{equation}
\begin{aligned}
  \hat{J}^{1342}_{\mu} &= \frac{Q_{214,3}}{s_{34}s_{234}}k^{1}_{\mu} 
  + \frac{1}{s_{34}} \bigg[ \frac{Q'_{1,432}}{ s_{134}} 
  +  \frac{(s_{24}+s_{34})(Q_{214,3} + Q'_{1,432})}{s_{24} s_{234}} \bigg] k^{2}_{\mu} 
  \\
  &\quad - \frac{Q_{234,1} + Q'_{1,234}}{s_{24} s_{234}} k^{4}_{\mu} + \bigg[ \frac{(s_{24}+s_{34}) P'_{3,421}}{s_{34} s_{234}} +  \frac{s_{14} P'_{3,412} }{s_{34} s_{134}} +\frac{P_{421,3}}{s_{234}}  \bigg] \epsilon^{4}_{\mu}
  \\
  &\quad + \bigg[ \frac{s_{12}\big(P_{124} (k_{4} {\cdot} \epsilon_{3}) {-} P_{123} (k_{3} {\cdot} \epsilon_{4})\big)}{s_{34} s_{234}} 
  {+}\frac{P_{1,243}}{ s_{234}} 
  +\frac{s_{12} Q_{432} {-} s_{14} Q_{234}}{2 s_{24} s_{234}}
  + \frac{s_{12} Q_{432}}{2 s_{34} s_{234}}\bigg] \epsilon^{1}_{\mu}
  \\
  & \quad
  + \bigg[ \frac{P_{2,341} {-} P_{2,431}}{ s_{134}} + \frac{P_{2,341} {-}P_{241,3} - P'_{3,241}}{s_{234}} + \frac{s_{12} P_{213,4} - s_{24} P_{241,3} }{s_{34} s_{234}} \bigg] \epsilon^{2}_{\mu}
  \\
  & \quad - \bigg[  \frac{(s_{24}+s_{34}) P_{3,421}}{s_{34} s_{234}} + \frac{s_{14} P_{3,412}}{s_{34} s_{134}} -\frac{s_{23} Q_{412} - (s_{24}+s_{34})Q_{214}}{2s_{24} s_{234}} \bigg] \epsilon^{3}_{\mu}
  \\
  & \quad +s_{1234} \big(\text{regular terms}\big) + k^{1234}_{\mu} \Lambda^{1342}
\end{aligned}
\end{equation}
where the regular terms are
\begin{equation}
\begin{aligned}
  -\frac{1}{4 s_{234}} \bigg[ \frac{Q_{432}-Q_{234}}{2s_{24}} + \frac{Q_{432}}{ s_{34} }\bigg]\epsilon^{1}_{\mu}\,,
\end{aligned}\label{}
\end{equation}
and
\begin{equation}
  \Lambda^{1342}=-\frac{1}{2} \bigg[ \frac{Q'_{1,432}}{s_{34} s_{134}}+ \frac{Q'_{1,432}-Q'_{1,234}}{4s_{24} s_{234}} +\frac{3Q_{214,3}+Q'_{1,432}}{2s_{34} s_{234}} \bigg]
\end{equation}
\\~\\
$5.~ \displaystyle \frac{s_{1234}}{2\sqrt{2}}  J^{1423}_{\mu}$
\begin{equation}
\begin{aligned}
  \tilde{J}^{1423}_{\mu} &= \bigg[ \frac{1}{s_{234}}{-}\frac{s_{13}+s_{34}}{s_{14} s_{124}}  \bigg] \frac{Q_{234,1}}{s_{23}} k^{1}_{\mu} 
- \bigg[ \frac{Q'_{1,432}}{s_{14}s_{124}} -\frac{(s_{13}-s_{24})(Q_{214,3} + Q'_{1,432})}{s_{24}s_{124}s_{234}} \bigg] k^{2}_{\mu}
\\
& \quad
- \bigg[ \frac{(s_{13}+s_{34})Q_{234,1}}{s_{14}s_{23}s_{124}} 
-\frac{(s_{13}s_{23} +s_{13} s_{24} + s_{24} s_{34})(Q_{234,1}+Q'_{1,234})}{s_{23}s_{24}s_{124}s_{234}}
\bigg] k^{4}_{\mu}
\\
& \quad
- \bigg[ \frac{P_{1,234}{+}P_{1,243}}{s_{234}} + \frac{P_{1,243}}{s_{124}}
- \frac{s_{12} (s_{13}-s_{24})S^{+}_{234}  }{2 s_{24} s_{124} s_{234}}
+\Big( \frac{s_{12}+s_{13}}{s_{23}}+\frac{s_{13}}{s_{24}} \Big) \frac{Q_{234}}{2 s_{234}} \bigg] \epsilon^{1}_{\mu}
  \\
  & \quad + \bigg[ \frac{P_{234,1}}{s_{14}} + \frac{ P'_{1,243} }{ s_{124}} + \frac{(s_{14}+s_{24})P_{243,1} }{ s_{14} s_{124}} +\frac{P'_{1,234}+P'_{1,243}+P_{234,1}+P_{243,1}}{s_{234}} \bigg] \epsilon^{2}_{\mu}
\\
& \quad + \bigg[ \frac{P_{3,412} + P_{3,421}}{s_{124}} + \frac{P_{3,421}}{s_{234}} 
  - \frac{ s_{34} (s_{13} - s_{24})S^{+}_{214}}{2 s_{24} s_{124} s_{234}} 
  + \Big( \frac{s_{13}+s_{34}}{s_{14}} +\frac{s_{13}}{s_{24}} \Big) \frac{Q_{412}}{2 s_{124}} \bigg] \epsilon^{3}_{\mu}
\\
& \quad
- \bigg[ \frac{P_{412,3}}{s_{23}} + \frac{P'_{3,421}}{s_{234}} +\frac{(s_{23}+s_{24})P_{421,3}}{s_{23} s_{234}} + \frac{P'_{3,412} +P'_{3,421} +P_{412,3} +P_{421,3}}{s_{124}}   \bigg] \epsilon^{4}_{\mu}
\\
& \quad
  +s_{1234} \big(\text{regular terms}\big) + k^{1234}_{\mu} \Lambda^{1423}\,,
\end{aligned}
\end{equation}
where the regular terms are
\begin{equation}
\begin{aligned}
  & \frac{Q_{214,3}}{s_{14}s_{23}s_{124}}k^{1}_{\mu}
  +\frac{(Q_{214,3} + Q'_{1,432})}{s_{24}s_{124}s_{234}}k^{2}_{\mu}
  +\frac{Q_{214,3}}{s_{14}s_{23}s_{124}}k^{3}_{\mu}
  +\frac{(s_{23} + s_{24})(Q_{214,3}+Q'_{1,234})}{s_{23}s_{24}s_{124}s_{234}}k^{4}_{\mu}
  \\
  & - \frac{1}{2s_{234}}\bigg[ \frac{s_{12} S^{+}_{234}  }{ s_{24} s_{124} } 
  -\frac{Q_{432}+3Q_{234}}{4s_{24} }
  -\frac{Q_{234}}{2s_{23}}
  \bigg] \epsilon^{1}_{\mu}
  \\
  &\quad
  + \frac{1}{2s_{124}}\bigg[\frac{  s_{34} S^{+}_{214}}{ s_{24}  s_{234}}
  -\frac{Q_{214}+3Q_{412}}{4s_{24} }
  -\frac{Q_{412}}{2s_{14}}\bigg] \epsilon^{3}_{\mu}
  - \frac{(Q'_{1,234}-Q'_{1,432})}{8s_{24} s_{124} s_{234}} k^{1234}_{\mu}\,,
\end{aligned}\label{}
\end{equation}
and
\begin{equation}
  \Lambda^{1423} = \frac{1}{2} \bigg[\frac{(s_{13} {-} s_{24})(Q'_{1,234} {-}Q'_{1,432})}{4s_{24} s_{124} s_{234}} - \frac{Q_{214,3}}{s_{14} s_{23}} - \frac{Q_{214,3}-Q'_{1,432}}{2s_{14} s_{124}} - \frac{3Q_{214,3}+Q'_{1,234}}{2s_{23} s_{234}} \bigg]\,.
\end{equation}
\\~\\
$6.~ \displaystyle \frac{s_{1234}}{2\sqrt{2}}  J^{1432}_{\mu}$
\begin{equation}
\begin{aligned}
  \tilde{J}^{1432}_{\mu} &= - \bigg[ \frac{1}{s_{34} }- \frac{s_{12}+s_{13}}{s_{14}s_{23}}  \bigg] \frac{Q_{214,3}}{s_{234}} k^{1}_{\mu}
  - \frac{1}{s_{23}} \bigg[ \frac{Q_{214,3}}{s_{14} } + \frac{Q_{214,3} + Q'_{1,234}}{ s_{234}} \bigg] k^{4}_{\mu}
  \\
  & \quad - \frac{1}{s_{34}}\bigg[  \frac{(s_{14}+s_{34})Q'_{1,432}}{s_{14}s_{134}}+\frac{Q_{214,3} + Q'_{1,432}}{s_{234}} \bigg] k^{2}_{\mu}
  \\
  & \quad
  + \bigg[ \frac{s_{12}(P_{123} k_{3} {\cdot} \epsilon_{4} - P_{124} k_{4} {\cdot} \epsilon_{3})}{s_{34}s_{234}} + \frac{P_{1,234}}{s_{234}} 
  + \frac{(s_{12}+s_{13})Q_{234}}{2s_{23}s_{234}}-\frac{s_{12}Q_{432}}{2s_{34} s_{234}}  \bigg] \epsilon^{1}_{\mu}
  \\
  & \quad
  - \bigg[  \frac{(s_{14}+s_{34})( P_{2,341} - P_{2,431}) }{s_{14}s_{134}} 
  - \frac{(s_{12}+s_{13})P_{234,1}}{s_{14} s_{234}}
  + \frac{P'_{1,234}+P_{2,341}}{s_{234}}
  \\
  & \qquad 
  +\frac{s_{12} P_{213,4} - s_{24} P_{241,3}  }{s_{34}s_{234}} \bigg] \epsilon^{2}_{\mu}
  + \bigg[  \frac{(s_{14}+s_{34})P_{3,412}}{s_{34} s_{134}} + \frac{s_{24}P_{3,421}}{s_{34} s_{234}} + \frac{Q_{412}}{2s_{14}} + \frac{S^{+}_{214}}{2s_{234}} \bigg] \epsilon^{3}_{\mu}
  \\
  & \quad
  - \bigg[  \frac{(s_{14}+s_{34})P'_{3,412}}{s_{34} s_{134}} - \frac{ P_{412,3} }{s_{23}} - \frac{s_{24}P_{421,3}}{s_{23} s_{234}}    + \frac{s_{24}P'_{3,421}}{s_{34}s_{234}}  \bigg] \epsilon^{4}_{\mu}
  \\
  & \quad
  +s_{1234} \big(\text{regular terms}\big) + k^{1234}_{\mu} \Lambda^{1432}\,,
\end{aligned}
\end{equation}
where the regular terms are
\begin{equation}
\begin{aligned}
   - \frac{Q_{214,3}}{s_{14}s_{23}s_{234}}k^{1}_{\mu}
+\frac{s_{23}Q_{432}-s_{34}Q_{234}}{4s_{23}s_{34}s_{234}}\epsilon^{1}_{\mu}
- \frac{P_{234,1}}{s_{14} s_{234}}\epsilon^{2}_{\mu}\,,
\end{aligned}\label{}
\end{equation}
and
\begin{equation}
  \Lambda^{1432}=\frac{1}{2} \bigg[ \frac{Q_{214,3}}{s_{14} s_{23}} + \frac{(s_{14}+s_{34})Q'_{1,432}}{ s_{14}s_{34}s_{134}} + \frac{3Q_{214,3}+Q'_{1,234}}{2s_{23} s_{234}} + \frac{3Q_{214,3}+Q'_{1,432}}{2s_{34} s_{234}} \bigg]\,.
\end{equation}


\section{ Gauge transformation terms in the rank-4 current KLT relation} \label{appendix_C}
In this appendix, we express the explicit form of $\Phi^{1234}_{\mu}$ in \eqref{gauge_transform_rank4}. It consists of two parts: terms proportional to $k^{i}_{\mu}$ and terms proportional to $\epsilon^{i}_{\mu}$
\begin{equation}
  \Phi_{\mu}^{1234} = A_{i} k^{i}_{\mu} + B_{i} \epsilon^{i}_{\mu}\,.
\label{}\end{equation}
Each $A_i$ term are as follows:
\begin{equation}
\begin{aligned}
  A_1 &= \frac{s_{123} s_{134}}{2 s_{24}} \left( \frac{s_{13} (s_{14}+s_{24})(s_{23} + s_{24}) - s_{12} s_{24} s_{34}}{s_{14} s_{23}} Q_{234,1}^{2} + 2 (s_{24} - s_{13}) Q'_{1,234} Q'_{3,412}
  \right) 
  \\
  A_2 &= \frac{Q'_{1,432} }{2 s_{24} s_{34}} \Bigg[\  2 Q_{432,1} s_{123} s_{134} (s_{23} s_{24} + s_{13} (s_{24} + s_{34}))   
  \\
  &\qquad\qquad\quad - \frac{Q'_{1,432}}{s_{14}} \bigg( s_{12} \Big(\big(s_{14} (s_{13} + s_{14}) (s_{13} + s_{14} + 2 s_{23}) +s_{13} s_{24} (2 s_{14} +s_{24})
  \\
  &\qquad\qquad\qquad\qquad\qquad\qquad + s_{24}(2 s_{14} +s_{23} +s_{24}) (s_{13} + 2s_{14} + s_{23}) + s_{24}^3 \big)  s_{34} 
  \\
  &\qquad\qquad\qquad\qquad\qquad\qquad + s_{14} s_{24} (s_{12} + s_{34})^2 - (s_{14} + s_{24}) s_{34}^2 (2 s_{12} + s_{34})\Big) 
  \\
  &\qquad\qquad\qquad\qquad\qquad  + (s_{13} + s_{23}) (s_{14} + s_{24}) \big(s_{13} s_{23} s_{34} - s_{12} (s_{14} + s_{34}) (s_{24} + s_{34})\big) 
   \\
  &\qquad\qquad\qquad\qquad\qquad +  s_{12}^2 \big(s_{234} s_{24} (s_{14} + s_{34}) + s_{14} s_{134} (s_{24} + s_{34})\big)\bigg) \  \Bigg] 
  \\
  A_3 &= -\frac{s_{123} s_{134} \left( s_{12} (s_{13}-s_{24}) (s_{24} + s_{34}) +s_{24} s_{123} s_{234} \right)}{2 s_{12} s_{24}s_{34}} Q_{214,3}^2
  \\
  A_{4} & = \frac{s_{134}}{2 s_{24}} \Bigg[\ \frac{2 (s_{13} (s_{23}+s_{24}) + s_{24} s_{34}) s_{123} Q'_{1,234} Q_{234,1}}{s_{23}}  
  \\
  &\qquad\qquad - (Q'_{1,234})^{2} \bigg( - \frac{s_{234} s_{24} s_{34}}{s_{12} } + \frac{s_{13} s_{24} s_{14}}{s_{23}} + s_{13} (s_{14} + s_{34})
+ \frac{s_{24} s_{34} (s_{14} -s_{23})}{s_{23}} \bigg) 
  \\
  &\qquad\qquad + (Q_{234,1})^2 s_{123}\bigg( \frac{s_{13}(s_{14}+s_{24})(s_{23} + s_{24}) - s_{12} s_{24} s_{34}}{s_{14} s_{23}}   
  \\
  &\qquad\qquad\qquad\qquad\qquad\quad + \frac{s_{23} (s_{13} - s_{24})}{s_{34}} -\frac{ s_{234} (s_{23} s_{24} + s_{13} (s_{12} + s_{24}))}{s_{12} s_{34}}\bigg)\ \Bigg]\,.
\end{aligned}\label{}
\end{equation}

Next we denote the $B_{i}$:
\begin{equation}
\begin{aligned}
  B_{1}&=\frac{Q_{234,1}}{2 s_{124} s_{234}} \bigg[ \Big(s_{13}+s_{14} + \frac{s_{13}(s_{14}+s_{24})-s_{12}s_{34}}{s_{23}}\Big) Q_{234}-\frac{(s_{13}+s_{14}) (s_{13}+s_{23})}{s_{34}} Q_{432}
  \\
  &\qquad\qquad\qquad + s_{12}s_{24}Q_{432}  + 2 s_{124} P_{1,234} 
 - \frac{ 2P_{124} (k_{4}{\cdot}\epsilon_{3})\big(s_{12} (s_{24}-s_{13}) -s_{123}s_{234}\big)}{  s_{34}} 
  \\
  &\qquad\qquad\qquad- 2P_{123} \Big((k_{3}{\cdot}\epsilon_{4}) \frac{(s_{13} + s_{14})  s_{124}}{s_{34}} -(k_{2}{\cdot}\epsilon_{4})  s_{234}\Big) \bigg] \epsilon^{1}_{\mu}
  \\
  B_{2} &= \frac{2P_{2,431} s_{34}}{s_{24} s_{124} s_{234}} \bigg[(s_{13}{-}s_{24})Q'_{1,234} + \frac{s_{12} s_{24} {+} s_{13} (s_{14} + s_{24})}{s_{14}}Q_{234,1} -\frac{s_{23} s_{24} + s_{13} (s_{24} {+} s_{34})}{s_{34}}Q_{432,1}
    \\
  &\qquad\qquad\quad -\frac{s_{13} s_{24} (s_{14} + s_{24}) s_{34} -
      s_{12} s_{23} s_{24} (s_{14} + s_{34}) + s_{12} s_{13} s_{14} (s_{24} + s_{34})}{ 2s_{14} s_{34} s_{134} } Q'_{1,432}  \bigg]
  \\
  & -\frac{ s_{12} s_{13} + (s_{13} + s_{14})s_{24}}{s_{24} s_{124} s_{234}} P_{214,3} Q'_{1,234}
  +\frac{ P_{213,4} (s_{13} +s_{14})-2 P_{2,341} s_{34}}{s_{34} s_{234}}Q_{432,1}
  \\
  & +  \frac{ s_{13} s_{34} + s_{12} (s_{14} + s_{34})}{s_{14} s_{134} s_{234}} P_{2,341} Q'_{1,432}
  - \bigg(\frac{(s_{13} - s_{24}) (s_{24} + s_{34})}{s_{24} s_{234}} +\frac{ s_{123} }{s_{12}}\bigg) \frac{ P_{214,3} Q_{432,1}}{s_{124} s_{34}}
  \\
  B_{3} &= 0
  \\
  B_{4}&= \frac{Q_{234,1}}{s_{124}} \bigg[ -\frac{s_{12} P_{4,213}}{s_{234}} \left( \frac{3 s_{13}}{s_{24}} +\frac{2 (s_{13}+s_{34})}{s_{23}} +\frac{s_{13}+s_{14}}{s_{12}} \right) +\frac{2 (s_{24}-s_{13})s_{34} P_{432,1}}{s_{24}s_{234}} 
  \\
  &\qquad\quad\ -\frac{ s_{34} P'{}_{1,432}}{s_{14} s_{24} s_{234}}  \left( s_{12}  s_{24}+s_{13} (s_{14} + s_{24})  \right)  -\frac{(s_{13} + s_{34}) s_{12} P_{4,123}}{s_{14} s_{23}} 
  \\
  &\qquad\quad\ +  \frac{(k_{2}{\cdot}\epsilon_{3}) (k_{4}{\cdot}\epsilon_{1}) (k_{4}{\cdot}\epsilon_{2})}{s_{14}  s_{234}}\left(\frac{s_{13} (s_{14} + s_{24}) - s_{12} s_{34}}{s_{23}} +s_{13}+ \frac{ s_{13}s_{14}}{s_{24}}\right) + \frac{s_{34} P_{431,2}}{s_{14}}
  \\
  &\qquad\quad\ + \frac{k_{1}{\cdot}\epsilon_{2}}{s_{12}}\Big(\frac{P_{213} s_{12} }{s_{12}s_{23}s_{123}}\left( s_{23} s_{34} +s_{12} (s_{13} + s_{34}) \right) - ke_{21} ke_{43}\Big)  \bigg]  
  \\
  & + \frac{Q'{}_{1,234} P_{4,213}}{s_{123} s_{124} s_{23} s_{234} s_{24}} \big(s_{12}s_{14}s_{24}(s_{13}+s_{34})+ s_{23}s_{24}s_{34}(s_{13}+s_{14}) +s_{12}s_{13}s_{23}(s_{14}+s_{34})\big)
\end{aligned}\label{}
\end{equation}
%

\newpage
\bibliography{references}
\bibliographystyle{JHEP}

\end{document}